\documentclass{silvia}

\usepackage{color} \usepackage{graphicx} \usepackage{subfigure}
\usepackage{txfonts}
\usepackage[normalem]{ulem}
\usepackage{amstext}

\begin{document}

   \title{Spectroscopically resolved far-IR observations of the massive star-forming region G5.89--0.39}

   \author{S. Leurini
          \inst{1}
          \and F. Wyrowski\inst{1} \and H. Wiesemeyer\inst{1}
          \and A. Gusdorf\inst{2,3}
          \and R. G\"usten\inst{1} \and K.\,M.\,Menten\inst{1}
          \and M. Gerin\inst{2,3}
          \and F. Levrier\inst{2,3}          
          \and H.\,W. H\"ubers\inst{4,5} 
          \and K. Jacobs\inst{6}
          \and O. Ricken\inst{1}
          \and H. Richter\inst{4}
   }

   \institute{Max-Planck-Institut f\"ur Radioastronomie, Auf dem H\"ugel 69, D-53121, Bonn, Germany\\\email sleurini@mpifr.de
     \and LERMA, Observatoire de Paris, \'Ecole Normale Sup\'erieure, PSL Research University, CNRS, UMR 8112, F-75014, Paris, France
     \and Sorbonne Universit\'es, UPMC Univ. Paris 6, UMR 8112, LERMA, F-75005, Paris, France
     \and Deutsches Zentrum für Luft-und Raumfahrt (DLR), Institute of Optical Sensor Systems, Rutherfordstrasse 2, D-12489, Berlin, Germany
     \and Humboldt-Universit\"at zu Berlin, Department of Physics, Newtonstr. 15, 12489 Berlin, Germany
\and K\"olner Observatorium f\"ur Submm Astronomie (KOSMA), I. Physikalisches Institut, Universit\"at zu K\"oln, Z\"ulpicher Str. 77, 50937 Cologne, Germany}

   \date{\today}

  \abstract
   {The fine-structure line of atomic oxygen at 63\,$\mu$m ([OI]$_{\rm{63\mu m}}$) is an
     important diagnostic tool in different fields of astrophysics: it is for example
     predicted to be the main coolant in several environments of
     star-forming regions (SFRs). However, our knowledge of this line relies
     on observations with low spectral resolution, and the real
     contribution of each component (photon-dominated region, jet) in the complex environment of SFRs  to
     its total flux is poorly understood.}
   {We investigate the contribution of jet and photon-dominated region emission, and of absorption to the
     [OI]$_{\rm{63\mu m}}$ line towards the hot gas around the ultra-compact H{\sc ii} region G5.89--0.39 and
     study the far-IR line luminosity of the source in different velocity regimes through
   spectroscopically resolved spectra of atomic oxygen, [CII], CO, OH, and H$_2$O.}
   {We mapped G5.89--0.39 in [OI]$_{\rm{63\mu m}}$ and in CO(16--15) with the GREAT receiver onboard SOFIA.
     We also observed the central position of the source in the ground-state OH \,$^2\Pi_{3/2}, J=5/2 \to J=3/2$ triplet and
     in the excited OH $^2\Pi_{1/2}, J=3/2 \to J=1/2$ triplets with SOFIA.
These data were complemented with APEX CO(6--5) and CO(7--6) maps and with {\it Herschel}/HIFI maps and single-pointing observations in lines of [CII], H$_2$O, and HF.}
   {The [OI] spectra in G5.89--0.39 are severely contaminated by
     absorptions from the source envelope and from different clouds along the
     line of sight. Emission is detected only at high velocities, and
     it is clearly associated with the compact north-south outflows
     traced by extremely high-velocity emission in low-$J$ CO
     lines. The mass-loss rate and the energetics  of
    the jet system derived from the [OI]$_{\rm{63\mu m}}$ line agree well with previous
     estimates from CO, thus suggesting that the molecular outflows in
     G5.89--0.39 are driven by the jet system seen in [OI]. The far-IR
     line luminosity of G5.89--0.39 is dominated by [OI] at
     high-velocities; the second coolant in this velocity regime is
     CO, while [CII], OH and H$_2$O are minor contributors to the total
     cooling in the outflowing gas. Finally, we derive abundances of
     different molecules in the outflow: water has low abundances
     relative to H$_2$ of $10^{-8}-10^{-6}$, and OH of
     $10^{-8}$. Interestingly, we find an abundance of HF to
     H$_2$ of $10^{-8}$, comparable with measurements in
     diffuse gas.}
   {Our study shows the importance of
spectroscopically resolved observations of the [OI]$_{\rm{63\mu m}}$ line
     for using this transition as diagnostic of star-forming regions. While this was not possible until now, the GREAT receiver onboard SOFIA has recently opened the possibility  of detailed studies of the
     [OI]$_{\rm{63\mu m}}$ line to investigate the potential of the transition for probing different environments.}


   \titlerunning{Atomic oxygen in the massive star-forming region G5.89--0.39}
   \maketitle
   

\section{Introduction}

The fine-structure line of atomic oxygen  at 63\,$\mu$m ([OI]$_{\rm{63\mu
          m}}$) has 
an important diagnostic value in several fields of astrophysics. It is expected
to be one of the main coolants in jets from young stellar objects
(YSOs; e.g., \citealt{1989ApJ...342..306H}) and therefore to be a direct
tracer of mass-loss rates \citep{1985Icar...61...36H}.
The [OI]$_{\rm{63\mu
          m}}$ transition is also predicted to be a major coolant in photon-dominated regions (PDRs)
\citep{1985ApJ...291..747T,1995ApJS...99..565S}, where, together with the [CII] fine structure line at 158$\mu$m, it can be used as diagnostic tool of the physical conditions \citep[e.g.,][]{1985ApJ...291..747T}. Observations show that  [OI]$_{\rm{63\mu m}}$ 
is an important PDR cooling line in external galaxies
      \citep[e.g.,][]{2001ApJ...561..766M,2004ApJ...604..565D,2012MNRAS.427..520C}.
      Since it suffers less from extinction than shorter wavelength lines, the [OI]$_{\rm{63\mu
          m}}$ line might also be a powerful tracer of
      star-formation rates in  galaxies even at high red-shifts
      \citep[e.g,][]{2014A&A...568A..62D}.  Studies of [OI]$_{\rm{63\mu m}}$  exist, mostly in
      galactic star-forming regions
      \citep{1996ApJ...462L..43P,1998ApJ...503..785K,2001ApJ...555...40G,2002ApJ...574..246N,1997ApJ...476..771C,2001ApJ...561..766M}.
      However, the diagnostic capabilities of this transition have not
      been fully exploited until now due to the poor angular and
      spectral resolutions and the poor sensitivity of previous
      instruments (ISO and KAO). This situation has recently
improved
with the PACS instrument onboard {\it Herschel}, which
      allowed observing atomic oxygen at 63\,$\mu$m with an angular
      resolution of 9\farcs4 and low spectral resolution
      \citep[e.g.,][]{2012A&A...545A..44P,2013ApJ...769L..13G,nisini2015}. However,
      previous observations and modelling of the [OI]$_{\rm{63\mu m}}$ line luminosity
      \citep[e.g., ][]{1996ApJ...462L..43P,2006A&A...446..561L,2015A&A...573A..30G,2015ApJ...801...72R} suggested 
      that spectrally unresolved observations of  the 63\,$\mu$m [OI] line can be difficult to interpret
      not only because several components (jet, PDRs) can contribute
      to its emission, but also because absorption from foreground
      clouds and self-absorption can contaminate the profile, thus
      undermining the diagnostic power of [OI]$_{\rm{63\mu m}}$ based on observations with
      poor spectral resolution. The GREAT\footnote{GREAT is a
        development by the MPI f\"ur Radioastronomie and the KOSMA/
        Universit\"at zu K\"oln, in cooperation with the MPI f\"ur
        Sonnensystemforschung and the DLR Institut f\"ur
        Planetenforschung.} instrument onboard SOFIA finally provides
      the astronomical community with the high spectral and angular resolution that is needed 
to      investigate the profile of the 63\,$\mu$m line and study its spatial
      distribution on a 6\farcs6 angular size scale.

\begin{figure*}[]
\centering
\includegraphics[width=18cm]{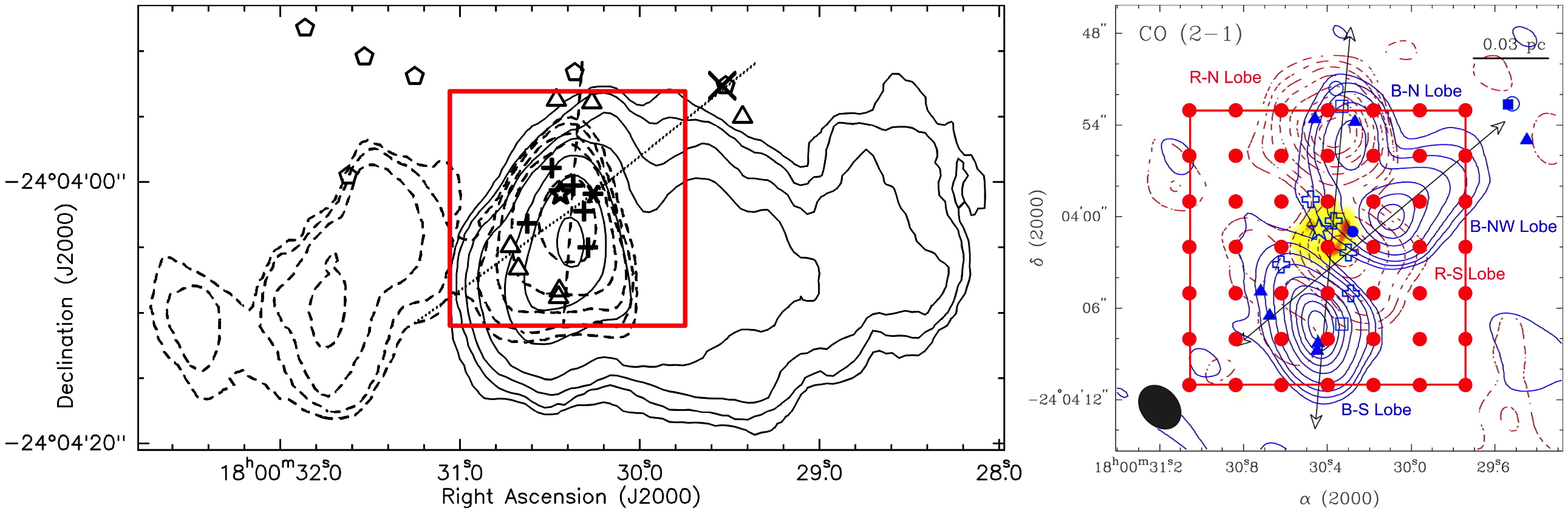}
\caption{Overview ({\bf left}) of the large scale  east-west  outflow and zoom ({\bf right})  in the inner region powering the extremely high-velocity molecular outflows in CO(2--1) imaged with SMA from \citet{2012ApJ...744L..26S}. The figures are adapted from \citet{2008ApJ...680.1271H} and \citet{2012ApJ...744L..26S}.
  In the left panel, contours show the large-scale  CO(1--0) outflow from \citet{2007ApJ...657..318W}, pentagons are class I methanol masers \citep{2004ApJS..155..149K}, while crosses
indicate  the
  submillimetre-millimetre  dust sources \citep{2008ApJ...680.1271H}.
  The dashed and dotted lines show the axes of the outflows detected at high velocities by \citet{2012ApJ...744L..26S}.
  In the right panel, the red box delineates the region mapped in [OI]$_{\rm{63\mu m}}$; the red dots are the centres of the [OI]$_{\rm{63\mu m}}$ raster map. 
  The overlaid colour scale represents the cm free–free emission \citep{2009ApJ...695.1399T},
  filled triangles are the H$_2$ knots, open circles are class I methanol maser positions, and open squares are positions of water masers \citep{1996A&AS..120..283H}. The crosses mark the position of the submillimetre-millimetre dust components. The ellipse shows the beam of the SMA observations from \citet{2012ApJ...744L..26S}.
}
\label{fig1}
\end{figure*}

We present here SOFIA observations of the massive star-forming region
G5.89--0.39 in [OI]$_{\rm{63\mu m}}$ aimed at resolving the distribution
of atomic oxygen in the source and studying its line profile.  The
observations are complemented by spectroscopically resolved data of
other main coolants (H$_2$O, OH, CO, [CII]) from SOFIA/GREAT, the High Frequency Heterodyne Instrument of the Far infrared (HIFI) onboard Herschel, and the Atacama Pathfinder Experiment 12 metre telescope (APEX)
to
study the contribution of each major contributor to the total far-IR
luminosity as a function of velocity.  G5.89--0.39 is an ideal target on which to
demonstrate the complexity that the [OI]$_{\rm{63\mu m}}$ emission can attain.  Its
distance ambiguity was recently solved through parallax measurements by
\citet{2011PASJ...63...31M}, who located it at 1.28\,kpc. It harbours
a shell-like ultra-compact H{\sc ii} region powered by a young O-type
star \citep[the so-called Feldt's star,][]{2003ApJ...599L..91F},
which is visible in
the near-IR. The region shows prominent outflow activity: at least
three outflows are associated with a hot dusty and molecular cocoon
surrounding the ultra-compact H{\sc ii} region
\citep{2008ApJ...680.1271H,2009ApJ...704L...5S}. The most prominent is
a gigantic outflow aligned along the east-west direction that
was discovered in CO emission by \citet{1988A&A...197L..19H}. The other two compact outflows
were detected with the Submillimter Array (SMA) by
\citet{2008ApJ...680.1271H} and  \citet{2012ApJ...744L..26S} and are 
associated with
extremely high-velocity material. These two flows are aligned along the
north-south and north/west-south/east directions and are associated
with H$_2$ knots \citep{2006ApJ...641..373P}.
The region is illustrated in Fig.\,\ref{fig1}, where the main
components of the source are labelled.

This paper is organised as follows: in Sect.\,\ref{sec_obs} we provide
the technical information related to the observations and the
calibration of the data performed with SOFIA and present the retrieved
{\it Herschel} and APEX archival data. In Sect.\,\ref{sec_res} we discuss
the [OI]$_{\rm{63\mu m}}$ profile and its morphology and compare the spectra of the
different atomic and molecular features toward the central position. In
Sect.\,\ref{sec_abu} the column densities and relative abundances of
different species are derived in two different velocity ranges from absorption and emission features.  In
Sect.\,\ref{sec_massloss} we derive the mass-loss rate  and other parameters describing 
the energetics of the outflow system from [OI]$_{\rm{63\mu m}}$. Finally, the total
far-IR line luminosity of the source is derived in
Sect.\,\ref{sec_fir} in different velocity ranges and over the total
line profile.

\section{Observations and data calibration}\label{sec_obs}
\begin{table*}
\caption[]{Summary of the observations. Frequencies are adapted
from the JPL and CDMS catalogues \citep{pickett_JMolSpectrosc_60_883_1998,2001A&A...370L..49M,2005JMoSt.742..215M}.}\label{obs}
\begin{center}
\begin{tabular}{lrcccccc}
 \hline  \hline
Line                                              &Frequency              &$\Delta \varv$&r.m.s. &Beam                       &$\eta_{\rm{MB}}$&Observing mode&Telescope\\
&\multicolumn{1}{c}{(GHz)}&(km\,s$^{-1}$)   &(K)      &\multicolumn{1}{c}{(\arcsec)}&\\
\hline
$[\rm{OI}]\,{^3P_1\to^3P_2}$                              &4744.7775              &0.5            &0.7   & 6.6                       &0.67        &map $18\arcsec \times 18\arcsec$&SOFIA\\
OH\,$^2\Pi_{1/2}, J=3/2, F=1^- \to J=1/2, F=1^+$   &1834.7350               &1.0&0.2&14.6                       &0.67          &single pointing&SOFIA\\
OH\,$^2\Pi_{1/2}, J=3/2, F=2^- \to J=1/2, F=1^+$   &1834.7469               &1.0&0.2&14.6                       &0.67          &single pointing&SOFIA\\
OH\,$^2\Pi_{1/2}, J=3/2, F=1^- \to J=1/2, F=0^+$   &1834.7499               &1.0&0.2&14.6                       &0.67          &single pointing&SOFIA\\
OH\,$^2\Pi_{1/2}, J=3/2, F=1^+ \to J=1/2, F=1^-$   &1837.7461               &1.0&0.2&14.6                       &0.67          &single pointing&SOFIA\\
OH\,$^2\Pi_{1/2}, J=3/2, F=2^+ \to J=1/2, F=1^-$   &1837.8163               &1.0&0.2&14.6                       &0.67          &single pointing&SOFIA\\
OH\,$^2\Pi_{1/2}, J=3/2, F=1^+ \to J=1/2, F=0^-$   &1837.8365               &1.0&0.2&14.6                       &0.67          &single pointing&SOFIA\\
OH\,$^2\Pi_{3/2}, J=5/2, F=2^- \to J=3/2, F=2^+$   &2514.2981               &1.5&0.6&11.6                       &0.70          &single pointing&SOFIA\\
OH\,$^2\Pi_{3/2}, J=5/2, F=3^- \to J=3/2, F=2^+$   &2514.3164               &1.5&0.6&11.6                       &0.70          &single pointing&SOFIA\\
OH\,$^2\Pi_{3/2}, J=5/2, F=2^- \to J=3/2, F=1^+$   &2514.3532              &1.5&0.6&11.6                       &0.70          &single pointing&SOFIA\\
CO\,$(6\to5)$                                    & 691.4730               &0.3&0.4& 9.0                     &0.52         &map $52\arcsec \times 52\arcsec$ \tablefootmark{a}&APEX\\
CO\,$(7\to6)$                                    & 806.6518               &0.5&0.8  & 7.7                   &0.49         &map $52\arcsec \times 52\arcsec$ \tablefootmark{a}&APEX\\
CO\,$(16\to15)$                                  &1841.3455               &2.0&0.4&14.5                       &0.65         &map $18\arcsec \times 18\arcsec$&SOFIA\\
$[\rm{CII}]\,^2P_{3/2}\to^2P_{1/2}$              &1900.5369               &1.0&0.8&11.2 \tablefootmark{b}          &0.59\tablefootmark{b}&map $46\arcsec \times 52\arcsec$ \tablefootmark{c}&{\it Herschel}\\
p-H$_2$O\,$ (2_{11}\to2_{02})$&752.0332&0.2&0.06&28.0\tablefootmark{b}              &0.64\tablefootmark{b}&single pointing\tablefootmark{d}&{\it Herschel}\\
p-H$_2$O\,$ (2_{02}\to1_{11})$&987.9268&0.2&0.1&21.2\tablefootmark{b}              &0.64\tablefootmark{b}&single pointing\tablefootmark{d}&{\it Herschel}\\
p-H$_2$O\,$(1_{11}\to0_{00})$                   &1113.3430               &0.5&0.2&18.9\tablefootmark{b}              &0.59\tablefootmark{b}&single pointing\tablefootmark{d}&{\it Herschel}\\
o-H$_2$O\,$(2_{21}\to2_{12})$                   &1661.0076               &0.1&0.4 &12.5\tablefootmark{b}             &0.55\tablefootmark{b}&single pointing\tablefootmark{d}&{\it Herschel}\\
o-H$_2$O\,$(2_{12}\to1_{01})$                   &1669.9048               &0.1&0.5 &12.5\tablefootmark{b}             &0.55\tablefootmark{b}&single pointing\tablefootmark{d}&{\it Herschel}\\
HF\,($1\to0$)                                    &1232.4763               &1.0&0.2&17.7\tablefootmark{b}            &0.59\tablefootmark{b}&single pointing         &{\it Herschel}\\
\hline
\end{tabular}
\end{center}
\tablefoot{\tablefoottext{a}{Observations presented by \citet{gusdorf15}.}
\tablefoottext{b}{{\it Herschel} HIFI beam sizes and efficiencies are based on the recent measurements from \citet{mueller14}}.}  
\tablefoottext{c}{Observations presented by \citet{2015A&A...573A..30G}.}
\tablefoottext{d}{Observations presented by \citet{2013A&A...554A..83V}.}

\end{table*}

\subsection{SOFIA observations}

The SOFIA observations we present here were carried out with the GREAT \citep{2012A&A...542L...1H} receiver during observatory cycles 1 and 2.
The ground-state transition of [OI] at 4744.77749~GHz was observed 
on 2014 May 17 (observatory cycle 2) with the H channel of GREAT using a novel waveguide hot electron bolometer heterodyne mixer with state-of-the-art sensitivity \citep{2015ITTST...5..207B}. The tuning of
its local oscillator, a quantum-cascade laser \citep{2013JIMTW..34..325H}, is fixed to the rest
frequency of this line. The Doppler correction was then applied off-line to the raw data.
The atmospheric opacity at this frequency is due to the wing
of a nearby water vapour feature and to the quasi-continuous collision-induced absorption
by N$_2$ and O$_2$. At the flight altitude of 13360\,m, the bulk of these
contributions is left below, and the residual water vapour along the sightline is lower
than 10~$\mu$m. The telluric [OI] line, originating from the mesosphere, contributes 
a significant but narrow absorption feature. We describe in the next
paragraph how we corrected for this. The resulting bandpass-averaged zenith opacity has a median value of 0.11; the double-sideband
system temperature was 2435~K. The CO $J=16 \rightarrow 15$ line at 1841.345~GHz was observed in
parallel with the L2 channel of GREAT with a system temperature (in the relevant part of the noise
bandpass) of 2012~K (single sideband) at a zenith opacity of 0.21. The observing mode for these
observations was a raster of seven by seven points, using the chopping secondary mirror 
of the telescope with alternating off-positions on either side of the observed sightline (with a chop
frequency and amplitude of 1~Hz and 60\arcsec, respectively). This mapping method yields the best
spectral baselines. The integration time per point was 40 seconds (comprising both chopper phases).

In cycle 1, G5.89--0.39 was observed on two southern deployment flights from New Zealand. On the first flight
(2013 July 24), the L1/L2 configuration of GREAT was installed. The L2 channel was tuned to the first
rotational line, $J=3/2 \rightarrow 1/2$, of the excited $^2\Pi_{1/2}$ OH state, at 1837.816~GHz (for
the exact frequencies of its hyperfine splitting see Table\,\ref{obs}). The median single-sideband system
temperature was 1705 K, for a zenith opacity of 0.1. We integrated for 12 minutes, with a chop frequency
and amplitude of 1~Hz and 30\arcsec, respectively. On the second flight  (2013 July 29), GREAT was
configured in the L2/M setup to simultaneously observe the first rotational line of the ground state of
OH, $^2\Pi_{3/2}, J=5/2 \rightarrow 3/2$ at 2514.317\,GHz (in the M channel) together with the excited OH
line, again for 12 minutes. Given the broad profile of the line, part of the second 
$^2\Pi_{3/2}, J=5/2 \rightarrow 3/2$ triplet at 2509.9\,GHz falls
in the observed band in the lower sideband of the receiver.
The median system temperature in the L2 channel is 1783~K, with a zenith
opacity of 0.07. The M channel had a system temperature of 4746~K and an opacity of 0.12.

The centre of the observations is the position $\alpha_ {\rm{[J2000]}} = 18^{\rm h}00^{\rm m}30\fs40$, $\delta_ {\rm{[J2000]}} = -24\degr04\arcmin02\farcs0$.

\subsubsection{Calibration and data reduction}\label{sec_cal}
The spectra were calibrated using loads at cold and ambient temperatures to 
determine the count-to-Kelvin conversion. Off-source sky measurements
then provided the total power of the atmospheric emission, from which the opacity
correction was derived using a dedicated atmospheric model, comprising the
aforementioned constituents of the atmospheric absorption. Details of the procedure
are given by \citet{2012A&A...542L...4G}. Our data were calibrated with the program {\tt kalibrate},
which is part of the {\tt KOSMA} software package. An accurate correction for the mesospheric
[OI]$_{\rm{63\mu}m}$ line based on existing atmospheric models  \citep[e.g., the AM model used in this work,][]{2014Paine...SMA...memo152} proved to be difficult. This is a consequence of the lack of high-resolution data
at this frequency in the past. We therefore adopted the following procedure: because of the high
altitude at which the line forms, its profile can be characterised by a Gaussian. 
For the opacity correction of our spectra, in which the telluric [OI]$_{\rm{63\mu}m}$ line appears in 
the outflow, we then adjusted the absorption strength  to achieve an adequate
interpolation between the adjacent unaffected spectral channels. A detailed description of the calibration steps for the [OI]$_{\rm{63\mu}m}$ line is given in Appendix\,\ref{cal}. The main-beam
efficiency ($\eta_{\rm mb} = 0.66$) was determined by means of observations
of planet Mars. The efficiency for the L2 channel, used in parallel for the CO line,
was 0.65. The efficiencies for the L2 and M channels flown on the southern deployment in 2013
were determined on Jupiter to 0.65 and 0.74, respectively. The flux scale accuracy was estimated to be about 20\%.

The calibrated spectra were further reduced (i.e., spectral averaging, baseline analysis, etc.)
with the {\tt CLASS90} software, which is a part of the {\tt GILDAS} software package, developed and maintained by IRAM. The CO(16--15) and the [OI]$_{\rm{63\mu m}}$ data were recentred on $\alpha_ {\rm{[J2000]}} = 18^{\rm h}00^{\rm m}30\fs40$,
$\delta_ {\rm{[J2000]}} = -24\degr04\arcmin00\farcs0$, which is the central position of mappings made with 
{\it Herschel} and APEX (see Sect.\,\ref{obs_apex_herschel}).
The [OI]$_{\rm{63\mu m}}$ map was produced with the {\tt XY\_MAP} task of {\tt CLASS90},
which convolves the data with a Gaussian of one third of the beam: the final angular resolution of the
[OI]$_{\rm{63\mu m}}$ data is 6\farcs96.

The angular and spectral resolutions of each dataset are reported in Table\,\ref{obs}. The 
  original spectral resolution is 0.005\,km\,s$^{-1}$ for the [OI]$_{63\mu\rm{m}}$ line, 0.02\,km\,s$^{-1}$ for CO(16--15), 0.01\,km\,s$^{-1}$ for the OH $^2\Pi_{1/2}, J=3/2 \to J=1/2$ triplets, and  0.005\,km\,$^{-1}$ for the OH $^2\Pi_{3/2}, J=5/2 \rightarrow 3/2$ ground-state triplet. Data were smoothed to the values reported in Table\,\ref{obs} to increase the signal-to-noise ratio.

\subsection{{\it Herschel} and APEX archive observations}\label{obs_apex_herschel}

The SOFIA observations presented in this paper are complemented by HIFI ([CII]\,$\rm{(^2P_{3/2}-^2P_{1/2}})$, p-H$_2$O\,$(2_{11}-2_{02})$, p-H$_2$O\,$(2_{02}-1_{11})$, p-H$_2$O\,$(1_{11}-0_{00})$, o-H$_2$O\,$(2_{21}-2_{12})$,   and o-H$_2$O\,$(2_{12}-1_{01})$ and HF(1--0)) and APEX (CO(6--5) and CO(7--6)) archival data.
The {\it Herschel}  \citep{2010A&A...518L...1P} observations
presented in this study were performed with the HIFI instrument \citep{2010A&A...518L...6D}.
All APEX \citep{2006A&A...454L..13G} and {\it Herschel} data are
centred at $\alpha_ {\rm{[J2000]}} = 18^{\rm h}00^{\rm m}30\fs40$,
$\delta_ {\rm{[J2000]}} = -24\degr04\arcmin00\farcs0$, thus at an offset of
$(0\arcsec,2\arcsec)$ from the centre of the SOFIA data.
We refer to \citet{2013A&A...554A..83V}, \citet{gusdorf15}, and \citet{2015A&A...573A..30G} for the details of the H$_2$O, [CII], and CO observations. 
The HF(1--0) data were acquired using the dual beam-switch observing mode in March, 2011. The data were processed with the standard HIFI pipeline
using {\tt HIPE}, and Level-2 data were exported using the
HiClass tool available in {\tt HIPE}. Further processing was performed
in {\tt CLASS90}. We assumed a flux scale accuracy of about 20\% for the HIFI observations\footnote{http://herschel.esac.esa.int/twiki/bin/view/Public/HifiCalibrationWeb\#\\HIFI\_performance\_and\_calibration} and for the APEX data \citep{gusdorf15}.
For the {\it Herschel} HIFI data (obtained in the framework of the Guaranteed Time Key Programs PRISMAS and WISH, PIs: M. Gerin and E. van Dishoeck, respectively), we used the new measurements of
\citet{mueller14} for the beam efficiencies and the beam sizes
for {\it Herschel} observations performed with the HIFI receiver (see Table\,\ref{obs}).

 The angular and spectral resolutions of each dataset are reported in Table\,\ref{obs}.  The 
  original spectral resolution is 0.08\,km\,s$^{-1}$ for the [CII] line, 0.15\,km\,s$^{-1}$ for p-H$_2$O\,$(2_{02}-1_{11})$, 0.13\,km\,s$^{-1}$ for p-H$_2$O\,$(1_{11}-0_{00})$, 0.09\,km\,s$^{-1}$ for o-H$_2$O\,$(2_{21}-2_{12})$ and $(2_{12}-1_{01})$, 0.3\,km\,s$^{-1}$ for HF, and 0.27\,km\,$^{-1}$ for CO(7--6). Data were smoothed to the values reported in Table\,\ref{obs} to increase the signal-to-noise ratio.

\section{Observational results}\label{sec_res}

\subsection{Atomic oxygen}\label{oi}

\begin{figure}[t]
\centering
\includegraphics[bb=0 90 782 520,clip,width=9cm]{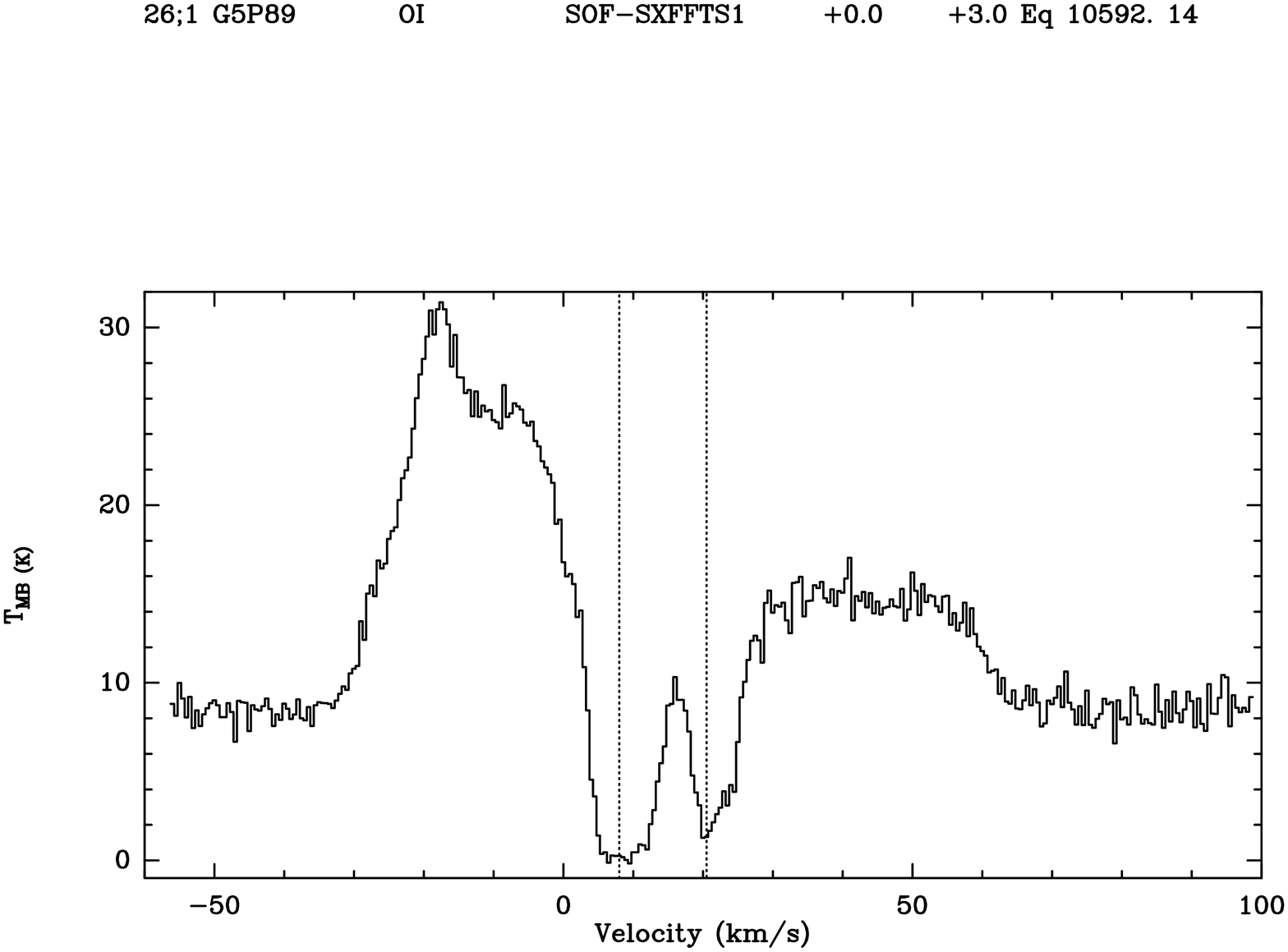}
\caption{[OI]$_{\rm{63\mu m}}$ spectrum at the original resolution (6\farcs6)
of the SOFIA data extracted from the strongest continuum position ($0\arcsec,+3\arcsec$ from $\alpha_ {\rm{[J2000]}} = 18^{\rm h}00^{\rm m}30\fs40$, $\delta_ {\rm{[J2000]}} = -24\degr04\arcmin02\farcs0$). The left vertical dotted line shows the absorption features at 2--15\,km\,s$^{-1}$  that are due to the source itself and to a cold cloud 
along the line of sight at $\varv_{\rm{LSR}}$ = 
13.7\,km\,s$^{-1}$. The right dotted vertical line shows an additional line-of-sight absorption feature.}\label{central}
\end{figure}

\begin{figure*}[t]
\centering
\subfigure[]{\includegraphics[bb = 77 157 527 593,clip,height=0.3\textheight]{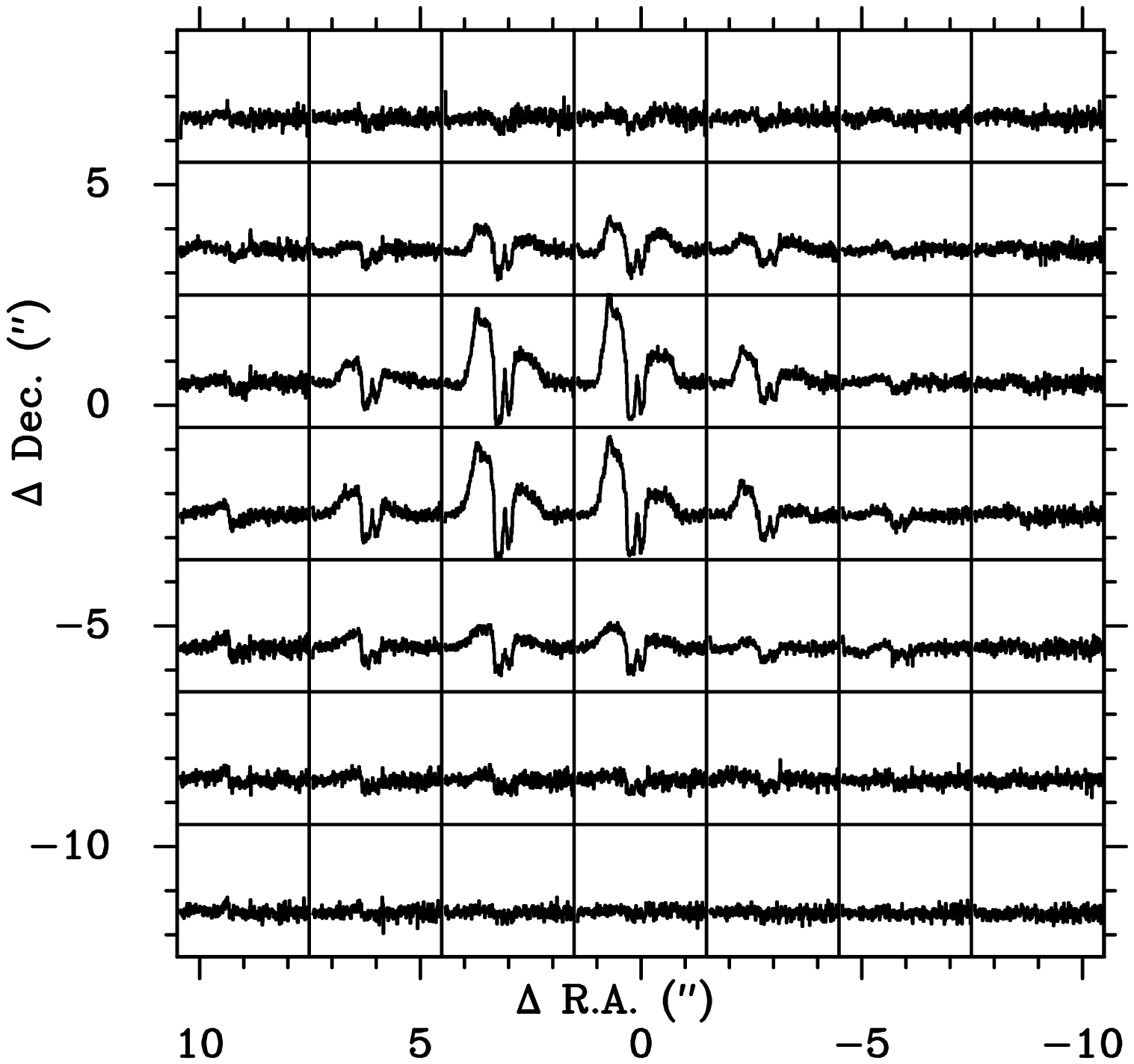}}
\subfigure[]{\includegraphics[bb = 0 103 595 650,clip,height=0.3\textheight]{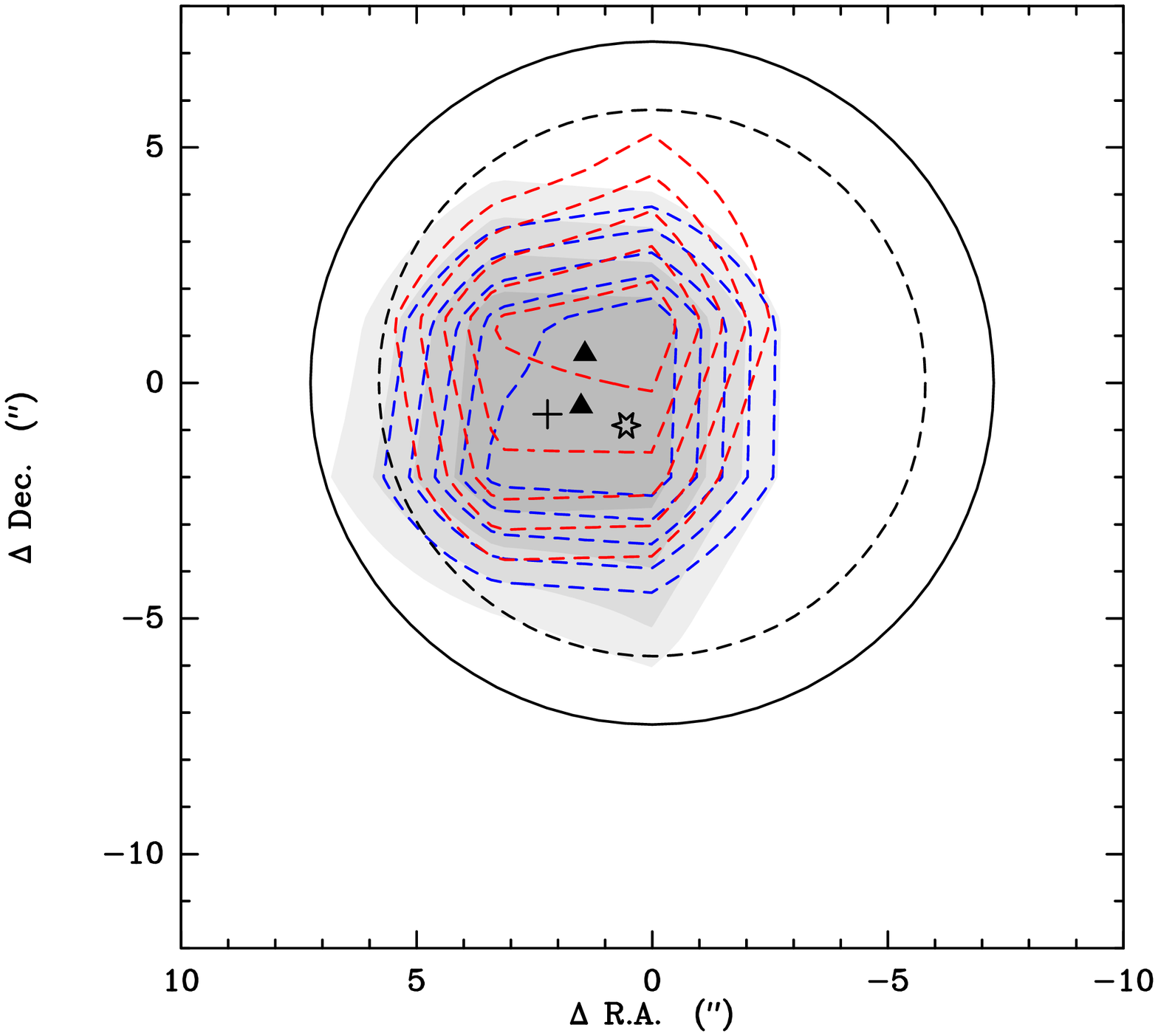}}
\caption{{\bf a:)} Continuum-subtracted spectral map of [OI]$_{\rm{63\mu m}}$ line in G5.89--0.39. The velocity range shown in the spectra ranges from $-$60\,km\,s$^{-1}$ to +100\,km\,s$^{-1}$. The temperature scale ranges from $-10$\,K to $+20$\,K.
  {\bf b:)} Map of the continuum emission at 63\,$\mu$m. Red and blue contours represent the integrated red- and blue-shifted intensity of the [OI]$_{\rm{63\mu m}}$ line. For the continuum emission and for the red- and blue-shifted integrated intensities levels are 50\,\% of the peak intensity (9.0\,K for the continuum, 187.5\,K\,km\,s$^{-1}$ for the red wing, 422.0\,K\,km\,s$^{-1}$ for the blue wing) in steps of 10\,\%.
  The star indicates the position of Feldt's star, the cross  the peak of the SOFIA 63\,$\mu$m continuum emission, the triangles are the peaks of the red- and blue-shifted [OI]$_{\rm{63\mu m}}$ emission (Table\,\ref{gaussianfit}). The solid and dashed circles represent the 14\farcs5 beam of the CO(16--15) data and the 11\farcs6 beam of the OH ground-state observations.
  In both panels the centre position is $\alpha_ {\rm{[J2000]}} = 18^{\rm h}00^{\rm m}30\fs40$, $\delta_ {\rm{[J2000]}} = -24\degr04\arcmin00\farcs0$.}\label{map}
\end{figure*}

Figure\,\ref{central} shows the [OI]$_{\rm{63\mu m}}$ spectrum at the original resolution (6\farcs6)
of the SOFIA data extracted from the strongest continuum position ($0\arcsec,+3\arcsec$ from $\alpha_ {\rm{[J2000]}} = 18^{\rm h}00^{\rm m}30\fs40$, $\delta_ {\rm{[J2000]}} = -24\degr04\arcmin02\farcs0$). The [OI]$_{\rm{63\mu m}}$ profile is characterised by emission at high
velocities (up to --40\,km\,s$^{-1}$ in the blue-shifted range, and
+67\,km\,s$^{-1}$ in the red-shifted wing) and strong absorption at
low velocities. The high-velocity emission covers a range very similar
to that of  other tracers, in particular, to that  of water.
We performed a
spatial 2D Gaussian fit on the integrated intensity red- and blue-shifted high-velocity 
emission maps in the velocity ranges [$+30,+67$]\,km\,s$^{-1}$) and
[$-40,-2$]\,km\,s$^{-1}$). The results are summarised in
Table\,\ref{gaussianfit}.
The high-velocity emission shows a similar distribution to the CO(6--5) emission
recently mapped by \citet{gusdorf15} with the APEX telescope with an angular resolution of 9\arcsec, but it is 
very compact compared to the CO(2--1) and
(3--2) extremely high-velocity emission mapped by
\citet{2012ApJ...744L..26S} with the SMA (Fig.\,\ref{fig1}). Both [OI]$_{\rm{63\mu m}}$
red- and blue-shifted emissions peak very close to the position of 
Feldt's star (see Fig.\,\ref{map}) with a relative offset between the
two lobes of 1\arcsec along the north--south direction, as also found
by \citet{gusdorf15} for CO(6--5). Thus, the high-velocity [OI]$_{\rm{63\mu m}}$
emission probably arises from the unresolved outflows mapped by
\citet{2012ApJ...744L..26S}. Since the bandwidth of the GREAT receiver at 63\,$\mu$m is limited to
the band shown in Fig.\,\ref{central}, our data cannot provide information on the possible presence of [OI]$_{63\mu\rm{m}}$ emission at extremely high velocities (up to --150\,km\,s$^{-1}$ at blue-shifted velocities and +90\,km\,s$^{-1}$ at red-shifted velocities) detected in CO(2--1) and (3--2) by \citet{2012ApJ...744L..26S}.

The deep absorption at
2--15\,km\,s$^{-1}$ is due to the source itself and to a cold cloud 
along the line of sight at $\varv_{\rm{LSR}}$ = 
13.7\,km\,s$^{-1}$ \citep[e.g.,][]{2006ApJ...648.1079K}, and it is completely
saturated (the single-sideband continuum main-beam brightness temperature is $9.0\pm0.7$\,K, see Table\,\ref{contlevel}). This suggests a large amount of [OI] with low excitation conditions in the source, since the outer envelope of 
the source is completely absorbed.
The features in absorption between $[+19,+25]$\,km\,s$^{-1}$ are
due to foreground clouds along the line of sight
\citep{2013ApJ...762...11F,2013A&A...554A..83V}.

\begin{table*}
\caption[]{Results of the 2D Gaussian fit of the [OI]$_{\rm{63\mu m}}$ integrated intensity and continuum distributions.}\label{gaussianfit}
\begin{center}
\begin{tabular}{lcccccc}
 \hline  \hline
Velocity range    &R.A. [J2000]&Dec. [J2000]&FWHM observed size&FWHM deconvolved size\\
\hline
-40/-2\,[km\,s$^{-1}$]& $18^{\rm h}00^{\rm m}30\fs51$ & $-24\degr04\arcmin00\farcs50$ &$8\farcs2\times 7\farcs0$ &$ 4\farcs3 \times 1\farcs1$\\
+30/+67\,[km\,s$^{-1}$]& $18^{\rm h}00^{\rm m}30\fs50$ & $-24\degr03\arcmin59\farcs38$ &$9\farcs1 \times6\farcs9$ &--\\
Continuum emission &$18^{\rm h}00^{\rm m}30\fs56$&$-24\degr04\arcmin0\farcs66$&$14\farcs2\times 12\farcs3$ &$ 12\farcs4 \times 10\farcs2$\\
\hline
\end{tabular}
\end{center}
\end{table*}

\begin{figure*}[t]
\centering
\includegraphics[width=18cm]{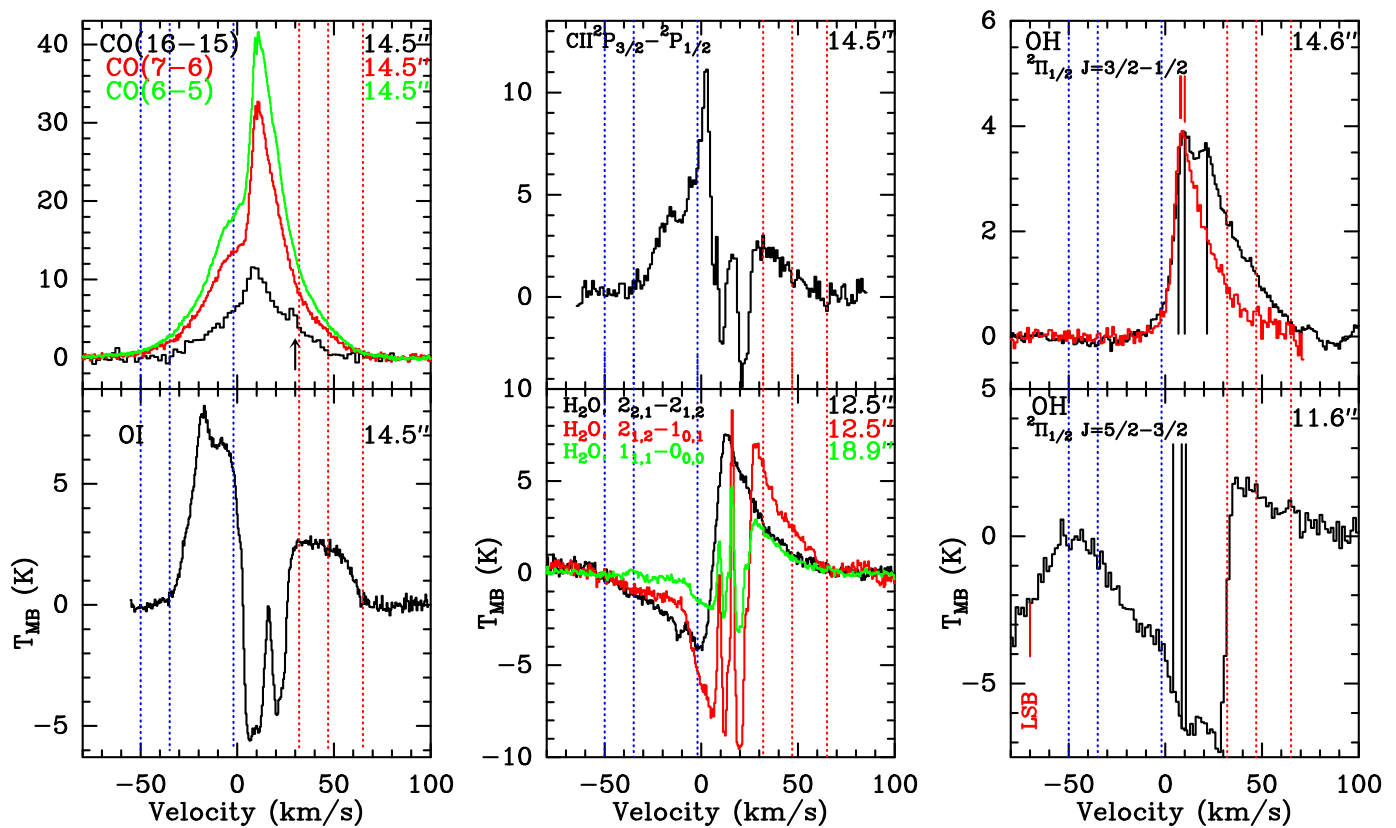}
\caption{Comparison of the [OI]$_{\rm{63\mu m}}$ line profile from the central position of the map 
with other molecular and atomic features. The [OI]$_{\rm{63\mu m}}$, [CII], CO(6--5) and CO(7--6) spectra are extracted from a map smoothed to  $14\farcs5$ to match
the resolution of the CO(16--15) data. The other lines are shown at their original  angular resolutions  (shown in each panel).
In all panels, the blue and red dashed lines indicate the HV and LV ranges (HV ranges: $\Delta_{\varv_{\rm{blue}}}=[-50,-35]$\,km\,s$^{-1}$;
$\Delta_{\varv_{\rm{red}}}=[+47,+65]$\,km\,s$^{-1}$; LV ranges: $\Delta_{\varv_{\rm{blue}}}=[-35,-2]$\,km\,s$^{-1}$;
$\Delta_{\varv_{\rm{red}}}=[+32,+47]$\,km\,s$^{-1}$). The arrow in the top left panel marks the feature at $+30$\,km\,s$^{-1}$
in the CO(16--15) spectrum due to a telluric line and probably also to a high-velocity bullet. The red label LSB in the bottom right panel with the OH $^2\Pi_{3/2}, J=5/2 \rightarrow 3/2$ triplet shows the blue-shifted
wing of the second triplet from the lower sideband of the receiver. The top right panel shows the two OH $^2\Pi_{1/2}, J=3/2\to1/2$ triplets at 1837\,GHz (black line) and 1834\,GHz (red line, detected in the lower sideband of the L2 GREAT receiver). The solid black and red vertical lines in the two panels with the OH lines show the three components of each triplet.}\label{spectra}
\end{figure*}

Figure\,\ref{spectra} shows the [OI]$_{\rm{63\mu m}}$ spectrum smoothed to an angular
resolution of 14\farcs5 (that of the GREAT CO(16–15) data,
Table\,\ref{obs}) and comparisons with CO, H$_2$O, OH, and [CII] lines smoothed to the same angular resolution where possible. The spectra are extracted at  $\alpha_ {\rm{[J2000]}} = 18^{\rm h}00^{\rm m}30\fs40$, $\delta_ {\rm{[J2000]}} = -24\degr04\arcmin02\farcs0$.

\subsection{CO(16--15)}\label{co1615}
\begin{figure}[t]
\centering
\includegraphics[width=8cm]{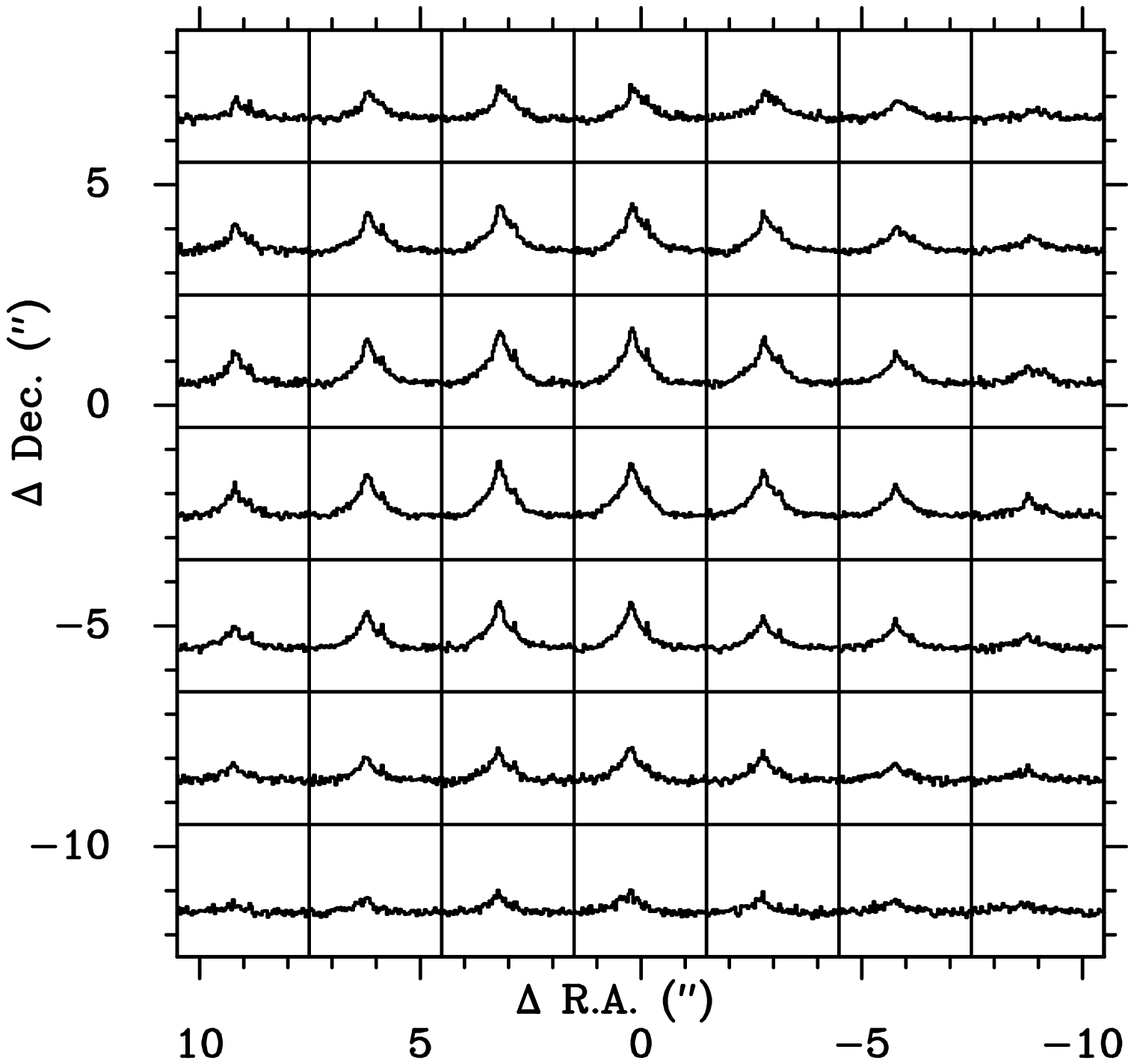}
\caption{Continuum subtracted spectral map of CO(16--15) in G5.89--0.39. Offsets are from the central position
$\alpha_{\rm{J2000}}=18^{\rm h}00^{\rm m}30\fs4$, $\delta_{\rm{J2000}}=-24\degr04\arcmin00\farcs0$. The velocity range shown in the spectra ranges from -60\,km\,s$^{-1}$ to +100\,km\,s$^{-1}$. The temperature scale ranges from $-10$\,K to $+20$\,K.
}\label{comap}
\end{figure}

The CO(16--15) spectrum towards the central position of our map is
shown in Fig.\,\ref{spectra}, while the spectral map is shown in
Fig.\,\ref{comap}. The red-shifted wing shows a narrow peak of emission at $+30$\,km\,s$^{-1}$; given the spatial extent of this feature (see Fig.\,\ref{comap}), this might be 
partially due to a 
telluric line  that is not completely removed by our calibration procedure (Sect.\,\ref{sec_cal}). However,
inspection of the HIFI CO(16--15) spectrum \citep[][
  Fig.\,14]{gusdorf15} reveals a weak peak at the same velocity, thus
suggesting that the $+30$\,km\,s$^{-1}$ feature in the SOFIA spectra is partially due to the telluric line and partially to a high-velocity peak.
Figure\,\ref{spectra} shows  comparisons
with the CO(6--5) and (7--6) 
smoothed to the same angular
resolution of CO(16--15).
High-velocity emission is detected in CO(16--15), although
in a narrower range than in the mid-$J$ CO lines, [OI]$_{\rm{63\mu m}}$, OH and
H$_2$O. This clearly indicates a decrease of excitation temperature with
velocity (see Sect.\,\ref{rotco}).  

\subsection{Hydroxyl radical}\label{oh}
Figure\,\ref{spectra} shows the spectra of the blended hydroxyl radical (OH) $^2\Pi_{1/2}, J=3/2\to1/2$ at 1834\,GHz and 1837\,GHz (detected in emission), and
$^2\Pi_{3/2}, J=5/2\to3/2$ lines (detected in absorption) at 2514\,GHz (see
Table\,\ref{obs} for the exact frequencies). The triplets at 1837\,GHz
and 2514\,GHz are partially resolved since they have a maximum
separation of 14.7\,km\,s$^{-1}$ and 6.6\,km\,s$^{-1}$; on the other
hand, the triplet at 1834\,GHz is unresolved since the lines are
separated by 2.4\,km\,s$^{-1}$ and the spectral resolution of this dataset is 1.5\,km\,s$^{-1}$ (see Table\,\ref{obs}).  The triplets at 1834\,GHz and 1837\,GHz have broad profiles with non-Gaussian red-shifted high-velocity emission in
a velocity range similar to that seen in [OI]$_{\rm{63\mu m}}$ and in the other
molecular species.  The hyperfine intensity ratios of the two triplets at 1834\,GHz and 1837\,GHz clearly deviate from the prediction of local thermal equilibrium (LTE) in the optically thin limit.
The triplet at 2514\,GHz is detected in absorption towards the
continuum of the source at rest velocity and at blue-shifted
high velocities. The foreground clouds at 20--25\,km\,s$^{-1}$ are also
detected in absorption. Weak red-shifted emission is detected up to $\sim+70$\,km\,s$^{-1}$ in agreement
with the profiles of the other molecular and atomic features. 
The spectrum at 2514\,GHz is complicated by
the detection of the blue-shifted wing of the outflow of the second
triplet at 2509\,GHz, which falls into the lower sideband of the GREAT
M channel  and is seen in Fig.\,\ref{spectra} at higher
negative velocities than the 2514\,GHz features. This
complicates the determination of the continuum level in the spectrum. If we assume that the blue-shifted
wing does not extend beyond $-40$\,km\,s$^{-1}$ (and indeed only [OI]$_{\rm{63\mu m}}$ and mid-$J$ CO are detected at such high
velocities) and that the red-shifted profile does not cover velocities greater than $+85$\,km\,s$^{-1}$
(not detected in any line except CO (2-1) and (3-2) by \citealt{2012ApJ...744L..26S}), we estimate a single-sideband continuum level of $6.7\pm0.8$\,K in $T_{\rm{MB}}$ units. The absorption around rest velocity and
around  $+25$\,km\,s$^{-1}$ is saturated.

\subsection{Hydrogen fluoride and ionised carbon}\label{obs_hf}

\begin{figure}[t]
\centering
\includegraphics[bb = 0 93 810 520,clip,width=9cm]{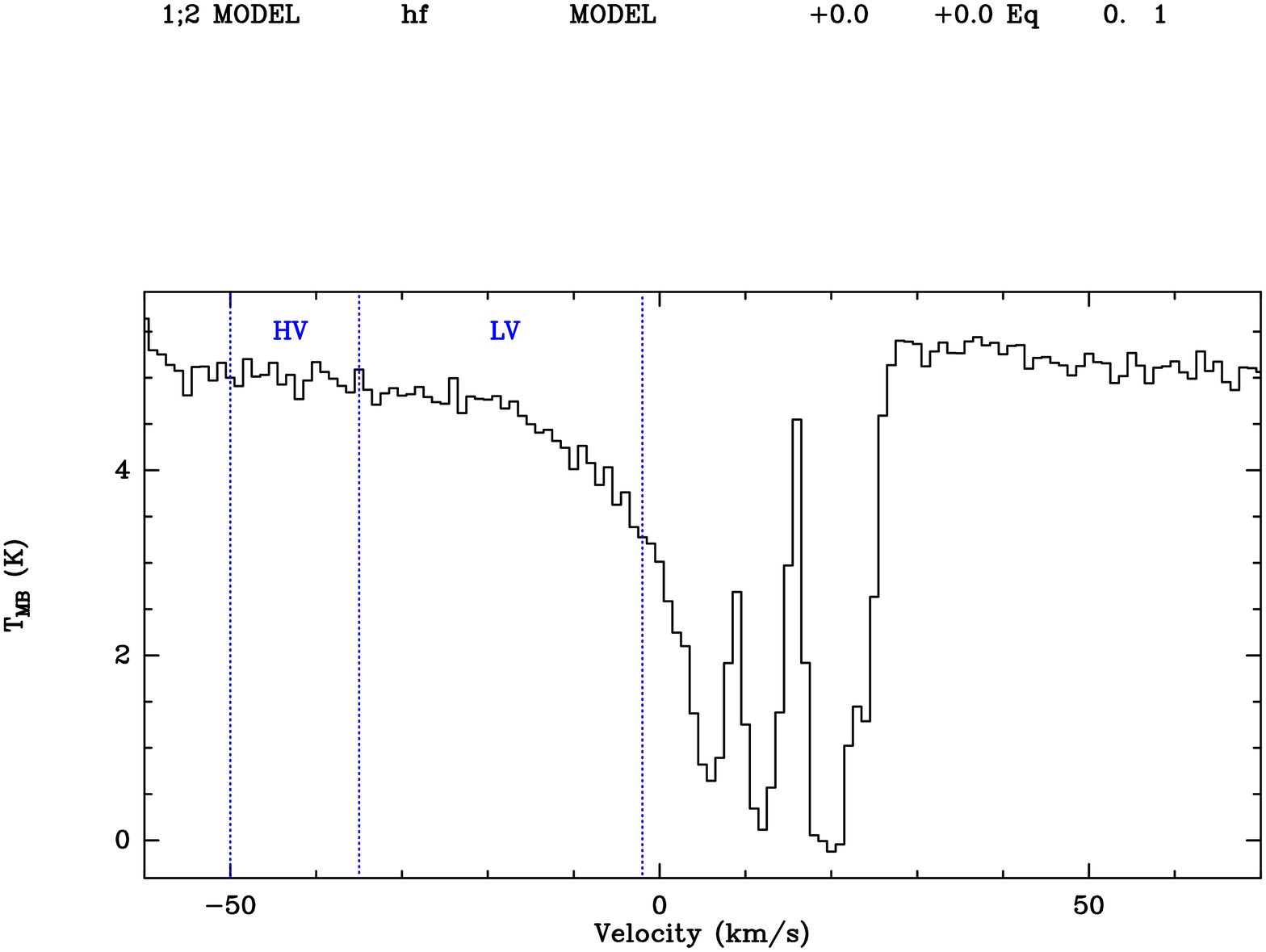}
\caption{HF(1--0) spectrum towards the position $\alpha_ {\rm{[J2000]}} = 18^{\rm h}00^{\rm m}30\fs40$, $\delta_ {\rm{[J2000]}} = -24\degr04\arcmin00\farcs0$. The vertical blue-dotted lines mark the low- ($-35,-2$\,km\,s$^{-1}$, LV) and high-velocity ($-50,-35$\,km\,s$^{-1}$, HV) regimes used to derive the abundances of several species in the blue-shifted wing. 
}\label{hf}
\end{figure}

The spectrum of the hydrogen fluoride (HF) (1--0) line of G5.89--0.39 is presented in Fig.\,\ref{hf}:
the line shows three narrow ($\Delta \varv\sim 3-7$\,km\,s$^{-1}$) absorption features centred at
$5.7$\,km\,s$^{-1}$,  $12.2$\,km\,s$^{-1}$, and $20.5$\,km\,s$^{-1}$, and a broad absorption 
at high velocities up to $\varv\sim-35$\,km\,s$^{-1}$. At red-shifted velocities weak emission is maybe detected up to $\sim+50$\,km\,s$^{-1}$, which could be caused by electron excitation, as suggested by \citet{2012A&A...537L..10V} for the Orion Bar. Given the tentative detection of emission in the red-shifted wing, we refrain from further analysis of HF at these velocities.
The single-sideband continuum level of G5.89--0.39 is $5.1\pm0.2$\,K in main-beam brightness temperature units, thus
the absorption features at $12.2$\,km\,s$^{-1}$, and $20.5$\,km\,s$^{-1}$ are saturated.

The ionised carbon ([CII]) fine structure $\rm{{^2P_{3/2}\to^2P_{1/2}}}$ line in G5.89--0.39 is presented in detail by
\citet{2015A&A...573A..30G} and \citet{gusdorf15}. It shows prominent high-velocity emission and absorption features associated with the source  and with foreground gas  at   $\sim+20$\,km\,s$^{-1}$ as in [OI]$_{\rm{63\mu m}}$, H$_2$O, and OH.  The continuum single-sideband brightness temperature is $10.5\pm0.8$\,K.

\subsection{Comparison of molecular and atomic spectra}

In Fig.\,\ref{spectra}, the [OI]$_{\rm{63\mu m}}$ spectrum from the central position of
the map is compared with other molecular (CO, H$_2$O, OH)
and atomic ([CII]) features observed at comparable angular
resolutions.  The spectra are convolved to an angular
  resolution of 14\farcs5 for the lines for which maps are available (see Table\,\ref{obs}). For the lines observed with single pointings, the angular resolution ranges from 11\farcs6 for the OH \,$^2\Pi_{3/2}, J=5/2 \to J=3/2$ triplet to 18\farcs9 for p-H$_2$O\,$(1_{11}\to0_{00})$.  In Fig.\,\ref{map}, we overlay the 14\farcs5 beam size (angular resolution of most of the data) and the 11\farcs6 beam of OH (the smallest) on the [OI]$_{63\mu{\rm{m}}}$ blue- and red-shifted integrated intensity map to show that the [OI]$_{63\mu{\rm{m}}}$ integrated intensity emission at 50\% of the peak intensity is fully covered by our smallest beam. For clarity, the beam size of each spectrum is also indicated in Fig.\,\ref{spectra}.

The profiles of all spectral features shown in Fig.\,\ref{spectra} are in all cases very similar and cover the
same velocity range.  The similarity of the line profiles strongly suggests that observations probe the same region despite the different angular resolutions.
Based on the comparison of the CO(16--15)
spectrum with the other profiles, we define four different velocity
ranges: the low-velocity (LV) blue- ($[-35,-2]$\,km\,s$^{-1}$) and
red-shifted ([$+32,+47]$\,km\,s$^{-1}$) wings, which are detected in
all spectral features (except for the blue-shifted LV in the OH $^2\Pi_{1_2}, J =3/2\to1/2$ triplets); the high-velocity (HV) blue-
($[-50,-35]$\,km\,s$^{-1}$) and red-shifted ([$+47,+65]$\,km\,s$^{-1}$)
ranges, which are detected in all lines except CO(16--15) (red- and blue-shifted HV) and OH and [CII] (blue-shifted HV). We define the red-shifted LV range starting from $+32$\,km\,s$^{-1}$ because the
CO(16--15) $[+28,+30]$\,km\,s$^{-1}$ channels are contaminated by a
narrow atmospheric feature (see Sect.\,\ref{co1615}).

Clearly, only lines with high critical densities (H$_2$O transitions,
HF(1--0), OH $^2\Pi_{3/2}, J = 5/2\to3/2$) are detected in absorption against
the continuum of the source at blue-shifted  velocities, while
transitions with low critical densities ([OI]$_{\rm{63\mu m}}$, CO lines, [CII]) are
detected in emission in the same velocity range. The OH $^2\Pi_{1/2}, J = 3/2\to1/2$ triplets, which
have an upper level energy comparable to CO(7--6) but a much higher
critical density, are the only lines not detected at blue-shifted high
velocities. 
The detection of the OH $^2\Pi_{1/2}, J = 3/2\to1/2$ triplets in
the red-shifted non-Gaussian wing  points to higher excitation
conditions than in the blue-shifted lobe.

\begin{figure}[t]
\centering
\includegraphics[width=9cm]{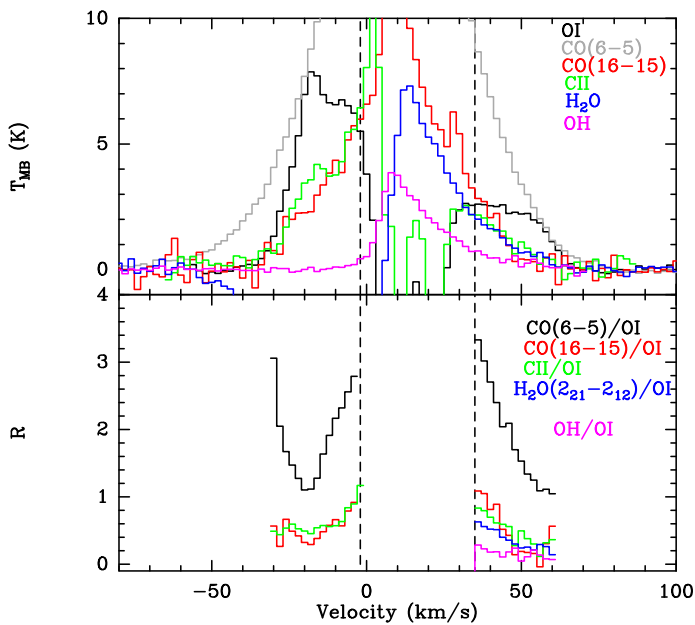}
\caption{{\bf Upper panel:} [OI]$_{\rm{63\mu m}}$ spectrum  from the central position of the map compared 
  with other molecular and atomic features. {\bf Lower panel:} Line ratio of CO(6--5) (black), CO(16--15) (red), [CII] (green), H$_2$O $2_{21}-2_{12}$ (blue), and the OH $^2\Pi_{1/2}, J=3/2\to1/2$ triplet  at 1834\,GHz (violet) to [OI]$_{\rm{63\mu m}}$. The dashed vertical lines show the velocities -2\,km\,s$^{-1}$ and +35\,km\,s$^{-1}$ inside which
the ratio is not computed.}\label{ratio}
\end{figure}

The similarities of the different line profiles are very well illustrated by Fig.\,\ref{ratio} where the line 
ratio of different transitions to the [OI]$_{\rm{63\mu m}}$ line is shown.  The ratio
is computed outside the velocity range
[--2,+35]\,km\,s$^{-1}$ and for the channels where the
[OI]$_{\rm{63\mu m}}$ line is above a 5\,$\sigma$ detection.  To
compute this ratio, we used the spectra shown in Fig.\,\ref{spectra} convolved to an angular
resolution of 14\farcs5 for those lines where maps are available. For OH, and H$_2$O, we
smoothed the [OI]$_{\rm{63\mu m}}$ data to the resolution of the other dataset. In both cases, we smoothed to a common
velocity resolution of 2\,km\,s$^{-1}$. Except for the CO(6--5)
lines, all other transitions have main-beam brightness temperatures lower than [OI]$_{\rm{63\mu m}}$
at high velocities; however, all line ratios behave similarly and
decrease with increasing velocity. The similarity of the profiles of OH and H$_2$O is
striking since in dissociative shocks  [OI]$_{\rm{63\mu m}}$ and OH are the main products of the destruction of water
\citep[e.g., ][]{1989ApJ...344..251N}. CO and H$_2$O lines
transitions behave similarly as a function of velocity: the  CO(6--5) and CO(16--15) to H$_2$O at 1661\,GHz line ratios are almost invariant
with velocity from $+15$\,km\,s$^{-1}$ (where the 1661\,GHz line is no longer affected by absorption) to $+45$\,km\,s$^{-1}$ (where the
CO(16--15) line intensity falls below 3$\sigma$). The CO(6--5) to
o-H$_2$O $(2_{21}\to2_{12})$ line ratio stays constant in the whole velocity range where the water
line is detected above 3$\sigma$.
These results differ from the findings
of increasing H$_2$O to CO line ratio with velocity in low-mass
Class 0 and Class I sources
based on the CO(3--2) and o-H$_2$O
ground-state lines \citep{2012A&A...542A...8K}, but fit the analysis of high-mass YSOs presented by \citet{sanjose_phd} for the CO(16--15) and (10--9) transitions, and the p-H$_2$O 987\,GHz line.
These findings suggest that mid- and high-$J$ CO lines ($J_{\rm{up}}>6$) and low-excitation H$_2$O transitions
originate in similar regions of the outflow system, probably in  cavity
walls \citep[e.g., ][]{2010A&A...518L.121V,2012A&A...537A..55V}.

\begin{table}
\caption[]{Single-sideband continuum levels.}\label{contlevel}
\begin{center}
\begin{tabular}{lcc}
 \hline  \hline
Frequency &$T_{\rm{MB}}$&Beam\\
\multicolumn{1}{c}{(GHz)}&\multicolumn{1}{c}{(K)}&\multicolumn{1}{c}{(\arcsec)}\\
\hline
4744.7775&$9.0\pm0.7$ & 6.6\\
2514.3531&$6.7\pm0.8$ &11.6\\
1900.5369&$10.5\pm0.8$\tablefootmark{a} &12.5\\
1661.0076&$10.0\pm0.5$ &12.5\\
1232.4763&$5.1\pm0.2$ &17.7\\
1113.3430&$3.8\pm0.2$ &18.8\\
\hline
\end{tabular}
\end{center}
\tablefoot{\tablefoottext{a}{\citet{2015A&A...573A..30G} report antenna temperature continuum levels.}}
\end{table}

\section{Column densities and abundances at high velocity}\label{sec_abu}

\subsection{Emission features}
\subsubsection{CO}\label{rotco}

\begin{figure}[]
\centering
\includegraphics[width=0.4\textwidth]{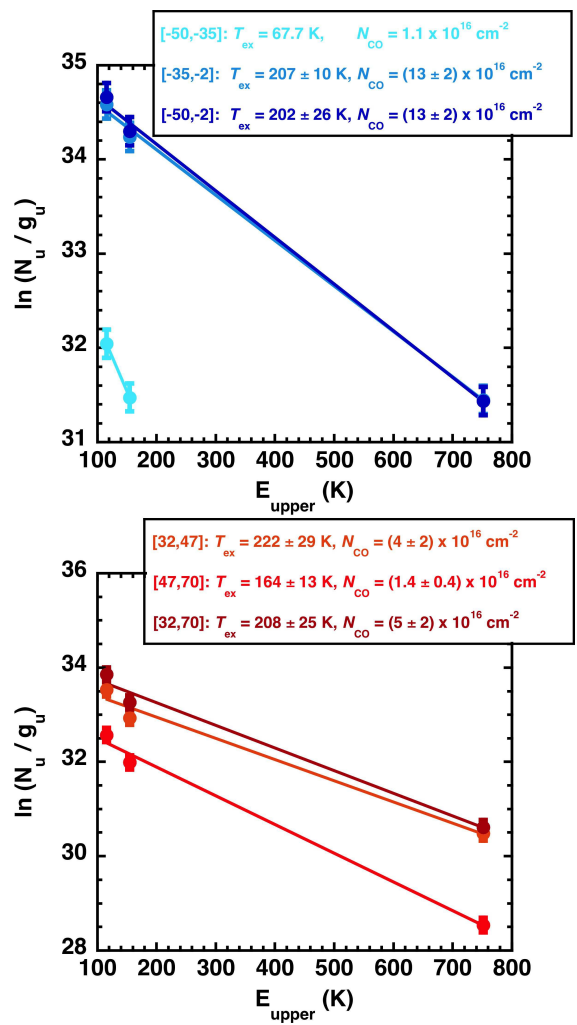}
\caption{CO rotational diagrams for different velocity ranges for the blue- (upper panel) and red-shifted
  (lower panel) emission
  over a beam size of 14\farcs5, with an assumed filling factor value of 1.}\label{corot}
\end{figure}

Using the CO(6--5), (7--6) and (16--15) lines, we generated 
rotation temperature diagrams \citep{1999ApJ...517..209G} in different velocity
ranges to determine the average excitation temperature and column
density of CO in a 14\farcs5 beam under the assumption that the emission from these lines
is optically thin. 
The results are presented in
Table\,\ref{boltzmann}  and Fig.\,\ref{corot}. There is a clear decrease of temperature
with increasing velocities especially in the blue-shifted wing where
the non-detection of CO(16--15) at HV constrains the excitation
temperature to $\le 68$\,K. Similarly, the red-shifted HV range
infers temperatures lower than in the LV red-wing. In
Table\,\ref{boltzmann} we also report  the values of $T_{\rm{ex}}$ and
$N_{\rm{CO}}$ over the whole blue-
  ($[-50,-2]$\,km\,s$^{-1}$) and red-shifted
  ($[+32,+70]$\,km\,s$^{-1}$) wings: these values agree well with the estimates for the warm component of \citet{gusdorf15}  through the analysis
of the whole CO ladder. Using the same method, these authors found a temperature of 270\,K
and a column density of $10^{17}$\,cm$^{-2}$ for both CO lobes by
fitting CO lines up to $J_{\rm{up}}=30$ from APEX and {\it
  Herschel}-HIFI and -PACS.

We note  that our result of decreasing temperature with
  velocity disagrees with the findings of
  \citet{2012ApJ...744L..26S}, who determined a clear trend
  of increasing temperature with gas velocity based on interferometric CO(2--1)
  and (3--2) observations. \citet{2012ApJ...744L..26S} performed their analysis on much higher angular resolution (3\farcs4 compared to our 14\farcs5), and this disagreement could be due to dilution over our larger beam.
  However, given the small energy range covered by
  these two lines ($\Delta {\rm{E_u}}\sim17$\,K), our dataset ($\Delta {\rm{E_u}}\sim600$\,K) is probably better suited to
  constrain the average temperature of the outflowing gas in the
  region as a function of velocity, although compact hot gas at high velocities could be diluted in our beam in the CO(16--15) data.

\begin{table}
\caption{Results of the rotational diagrams for CO.}\label{boltzmann}
\begin{center}
\begin{tabular}{lcc}
\hline  \hline
Velocity range &$T_{\rm{ex}}$&$N_{\rm{CO}}$\\
& (K) & $(10^{16}$\,cm$^{-2})$\\
\hline
HV-blue  ($[-50,-35]$\,km\,s$^{-1}$)  & 67.7 &1.1\\
LV-blue  ($[-35,-2]$\,km\,s$^{-1}$)    &$207\pm10$&$13\pm2$\\
blue-wing ($[-50,-2]$\,km\,s$^{-1}$)   &$202\pm26$&$13\pm2$\\
HV-red  ($[+47,+70]$\,km\,s$^{-1}$)   &$164\pm13$ &$1.4\pm0.4$\\
LV-red  ($[+32,+47]$\,km\,s$^{-1}$)   &$222\pm29$ &$4\pm2$\\
red-wing ($[+32,+70]$\,km\,s$^{-1}$)  &$208\pm25$ &$5\pm2$\\

\hline
\end{tabular}
\end{center}
\end{table}

\subsubsection{Atomic oxygen}\label{col_oi}

Given the complexity of the [OI]$_{\rm{63\mu m}}$ profile, we cannot
use the PACS [OI]$_{145\,\mu{\rm m}}$ flux \citep{2014A&A...562A..45K} to
study the excitation of [OI] through the analysis of the
[OI]$_{63\,\mu{\rm m}}$ to [OI]$_{145\,\mu{\rm m}}$ line ratio as done by other authors
\citep[e.g., ][]{nisini2015}. We therefore estimated the [OI] column density, $N_{\rm{[OI]}}$, from the integrated intensity of the 63\,$\mu$m line
  using the LTE approximation. Under this assumption, $N_{\rm{[OI]}}$ is

\begin{equation}\label{eq_coloi}
  N_{\rm{[OI]}}=\frac {8\pi k \nu^2}{h c^3 A_{ul}}\times\frac{Q_{T_{\rm{ex}}}exp(E_u/kT_{ex})}{g_u}\int{T_{\rm{MB}}d\varv}\frac{\tau}{1-exp(-\tau)}
,\end{equation}

where $Q_{T_{\rm{ex}}}$ is the partition function, $T_{\rm{ex}}$ is
the excitation temperature, $g_u$ is the statistical weight of the
upper level of the transition, $A_{ul}$ is the Einstein coefficient for spontaneous emission,
and $\tau$ the optical depth of the 63\,$\mu$m line. 
For
the blue-shifted wing, assuming an excitation temperature of 200\,K (as derived for
CO)
and without corrections for dilution, we derive an averaged optical
depth of 0.06.  This corresponds to an average column density over the
6\farcs6 beam of $\sim 1.2\times10^{18}$\,cm$^{-2}$. For the red-shifted wing,
the average column density over the
6\farcs6 beam is $\sim 5.7\times10^{17}$\,cm$^{-2}$ with $\tau=0.02$ and $T_{\rm{ex}}=200$\,K.
For a direct comparison with estimates of column densities of CO, OH,
and H$_2$, which are detected on larger scales, we also computed the
average column density of [OI] on a 14\farcs5 beam,
$N_{\rm{[OI]}_{14\farcs5}}= 5.3\times10^{17}$\,cm$^{-2}$ for the blue-shifted emission and $N_{\rm{[OI]}_{14\farcs5}}=2.1\times10^{17}$\,cm$^{-2}$ for the red-shifted wing. Finally,
we also report a [OI] column density of
$N_{\rm{[OI]}_{12\farcs5}}=3.2\times10^{18}$\,cm$^{-2}$ for the
blue-shifted wing and of $\sim 2.9\times10^{18}$\,cm$^{-2}$ for the
red-shifted wing. These values are integrated over a beam of 12\farcs5 and corrected
for a source size of 6\farcs6. These estimates are used in
Sect.\,\ref{sec_massloss} to derive the energetics parameters of the
jet system and compare with similar estimates for the molecular
outflows inferred by \citet{gusdorf15} on a 12\farcs5 scale.

We note, however, that the red-shifted
flat profile suggests that the emission at these velocities has a very
high optical depth since the profile is almost flat. Therefore, the column densities derived for the high-velocity red-shifted
emission under the assumption of optically thin emission are lower limits to the real values. 
Under the  assumption of extremely high optical depth, we  can estimate the size of the red-shifted lobe
assuming $T_{\rm{ex}}=200$\,K.  The red-shifted observed main-beam
brightness temperature corrected for the fact that the Rayleigh-Jeans
approximation is not valid in this frequency and temperature regime is
$\sim 63$\,K. This implies a circular size of diameter 4\farcs5 for
the red lobe, in agreement with the findings of Sect.\,\ref{oi}.

\subsubsection{Ionised carbon}\label{cii}
We determined the column density of [CII], $N_{\rm{[CII]}}$, in the blue- and
red-shifted wings  based on the LTE
assumption as done for [OI]$_{\rm{63\mu m}}$. Using Eq.\,\ref{eq_coloi} adapted for [CII],
we obtain a [CII]  column
density of $3.6\times10^{18}$\,cm$^{-2}$ in the blue-shifted wing of
the spectrum (between -50 and -2\,km\,s$^{-1}$), and $1.2\times10^{18}$\,cm$^{-2}$
in the red-shifted wing of the
spectrum ([+32,+70]\,km\,s$^{-1}$). These values  are obtained over a beam size of 14\farcs5 assuming optically thin emission,
an excitation temperature of 200\,K, and a source size of 12\farcs5.

As already discussed by \citet{gusdorf15}, [CII] is the dominant form of carbon in the warm outflowing gas.

\subsubsection{H$_2$O and OH}\label{col_radex}

To determine the column density of OH and H$_2$O at red-shifted
velocities, we modelled their emission with the RADEX program
\citep{2007A&A...468..627V}, a radiative transfer code based on the
large velocity gradient (LVG) approximation with a plane-parallel slab
geometry. We investigated models with column densities ranging from
$10^{13}$\,cm$^{-2}$ to $10^{18}$\,cm$^{-2}$, kinetic temperatures
$25\,{\rm{K}} \le T \le375\,{\rm{K}}$, and densities in the range
$10^3-10^8$\,cm$^{-3}$. Ortho and para water were analysed separately and no assumptions on their relative abundance was made in the calculations.
We considered a line width of 38\,km\,s$^{-1}$
(corresponding to the total velocity range $[+32,+70]$\,km\,s$^{-1}$ of the red-shifted emission),
and a source
size of 12\farcs5 for all lines. This is based on the CO source size
considered by \citet{gusdorf15}.  The molecular input files are
taken from the
LAMDA database \citep{2005A&A...432..369S}; for OH, we used entries
that did not take into account the hyperfine structure because of the
severe line overlap of the dataset.  For both molecules, two solutions
are found with similar values of the reduced $\chi^2$: the first gives
column densities of some $10^{14}$\,cm$^{-2}$, high densities
($\ge10^7$\,cm$^{-3}$), and temperatures higher than 100\,K; the
second solution implies high column densities ($\sim
10^{17}$\,cm$^{-2}$), low densities ($n\le10^6$\,cm$^{-3}$), and no
constraints on $T$. In neither case do the models fit the
observations well for OH and p-H$_2$O.
This is most likely due to the
simplifying assumptions made in the analysis (uniform temperature, density, and
column density; common source size).
Since the first
solution matches the physical parameters derived from CO
(see Sect.\,\ref{rotco}) fairly well, we chose this as the best
fit to our data.

\subsection{Absorption features}\label{abs}

 Since the blue-shifted wing of several molecular lines is detected in absorption against the corresponding continuum of G5.89--0.39, we can estimate the column density of OH, H$_2$O, and HF at these velocities 
through the relation

\begin{equation}\label{tau}
\tau=-\mathrm{ln}\left(\frac{T_{L}-J_\nu(T_{\rm{ex}})\Omega_s}{T_{c}\Omega_c-J_\nu(T_{\rm{ex}})\Omega_s}\right)
,\end{equation}

 where $T_{L}$ and $T_{C}$ are the brightness temperatures of the absorption line and of the continuum, $J_\nu(T_{\rm{ex}})$ is the Rayleigh-Jeans equivalent radiation field due to the molecular emission with excitation temperature $T_{\rm{ex}}$, $\Omega_c$ and $\Omega_s$ are the dilution factors for the continuum source and the blue-shifted outflow lobe, respectively. For the continuum we adopted the size derived at 63\,$\mu$m ($\sim12\arcsec$, see Table\,\ref{gaussianfit}) at the other frequencies, and we used a similar size (12\farcs5) for the blue-shifted lobe as inferred from CO. From Eq.\,\ref{tau} we can also estimate an upper limit to the excitation temperature of the lines seen in absorption. We derive an upper limit
  of $\sim30$\,K and 25\,K to the excitation temperatures of the 1661\,GHz and 1669\,GHz water lines, and of 14\,K for the HF ($1\to0$) transition. For H$_2$O, we cannot use the $1_{11}\to0_{00}$ line at 1113\,GHz because the line-to-continuum ratio in
the blue-wing can only be reliably computed in a few channels. Assuming LTE conditions, the column density of the upper ($N_u$) and
lower level ($N_l$) are related to the total column density
$N_{\rm{tot}}$ of a molecule through
$N_{\rm{tot}}=Q_{\rm{T_{ex}}}\times\frac{N_u}{g_u}\exp(\frac{E_u}{k{\rm{T_{ex}}}})$.
The column density of a given molecule at a given
excitation temperature can be derived in the same way as for [OI] (see Eq.\,\ref{eq_coloi}, Sect.\,\ref{col_oi}). Results are listed in Table\,\ref{abu}. For the 1669\,GHz line we used a lower excitation temperature (12\,K) than the upper limit of 25\,K because the column density at 25\,K is too high compared to the value obtained from the 1661\,GHz transition.

\begin{table}
\caption{Column densities of different species and abundances relative to CO.}\label{abu}
\begin{center}
\begin{tabular}{lccc}
  \hline  \hline
\multicolumn{4}{c}{Blue-shifted wing $(\Delta{\varv}=[-50,-2]\,{\rm{km\,s^{-1}}})$}\\
Species&$N_{X_{\rm{Tex}}}$&$X=\frac{N_{X_{\rm{Tex}}}}{N_{\rm{CO}}}$&T$_{\rm{ex}}$ \\
&(cm$^{-2}$)&&(K)\\
\hline
HF              &$(2.7\pm0.5)\times10^{14}$&$2\times10^{-3}$&14\\
CO\tablefootmark{a}    &$(1.3\pm0.4)\times10^{17}$&&202\\
CII           &$(3.6\pm0.8)\times 10^{18}$  &28&200 \\
OH              &$(3.5\pm0.1)\times10^{14}$&$3\times10^{-3}$&10\\
o-H$_2$O (1661\,GHz)&$(3.0\pm0.1)\times10^{16}$&$2\times 10^{-1}$&30\\
o-H$_2$O (1669\,GHz)&$(5\pm0.3)\times10^{15}$&$4\times10^{-2}$&10\\

\hline
\multicolumn{4}{c}{Red-shifted wing $(\Delta{\varv}=[+32,+70]\,{\rm{km\,s^{-1}}})$}\\
\hline
CO\tablefootmark{a}      &$(5\pm2)\times10^{16}$&&208\\
CII           &$(1.2\pm0.7)\times 10^{18}$ &24 &200 \\
OH\tablefootmark{b}      &$3\times10^{14}$&$6\times10^{-3}$&>100\\
o-H$_2$O\tablefootmark{b}&$3\times10^{14}$&$6\times 10^{-3}$&>100\\
p-H$_2$O\tablefootmark{b}&$2\times10^{14}$&$4\times 10^{-3}$&>100\\

\hline
\end{tabular}
\end{center}
\tablefoot{
\tablefoottext{a}{From the rotational diagram analysis (Table\,\ref{boltzmann}).}
\tablefoottext{b}{From the RADEX-LVG analysis (Sect.\,\ref{col_radex}). We do not report error bars in the estimates of $N_{\rm{OH}}$ and  $N_{\rm{H_2O}}$ due to the large uncertainties.}}
\end{table}

\begin{figure}[t]
\centering
\includegraphics[width=8cm]{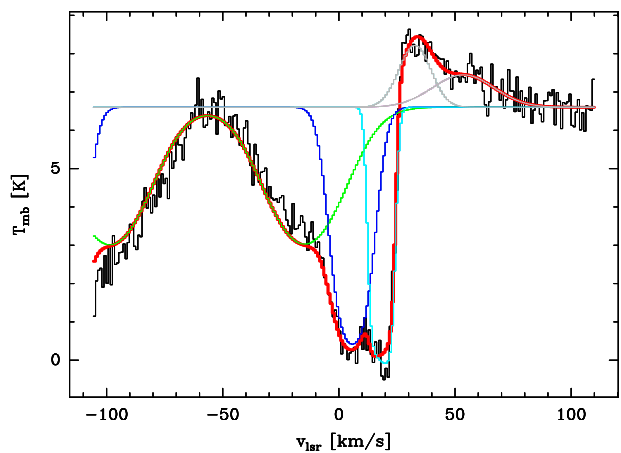}
\caption{Spectrum of the OH $^2\Pi_{3/2}, J=5/2 \rightarrow 3/2$ triplet at 2514\,GHz. The absorption feature at
  $\varv<-60$\,km\,s$^{-1}$ is the blue-shifted wing of the second $^2\Pi_{3/2}, J=5/2 \rightarrow 3/2$ triplet at 2509\,GHz observed in the lower sideband of the receiver. The red solid line represents the best fit and it  is given by the sum of the spectra of all velocity components used in Sect.\,\ref{abs}; green, blue, and cyan
lines are the synthetic spectra of the absorption features
at -13.2\,km\,s$^{-1}$, +7.3\,km\,s$^{-1}$ , and +20.2\,km\,s$^{-1}$, respectively;
the two grey lines represent the spectra of the emission components
at +34.3\,km\,s$^{-1}$ and 55.1\,km\,s$^{-1}$.}\label{oh_fit}
\end{figure}

For the OH ground-state line,
the hyperfine structure was modelled following Wiesemeyer et al. (2015, subm.), using the appropriate
weights for the opacity of each component. Five velocity components
were used to fit the line profile: one to fit the absorption at blue-shifted velocities from the
outflow, two for the  absorptions from
different lines of sights and from the source envelope, and two to fit the red-shifted wing we see in emission.
We used an excitation temperature of 10\,K for the absorption features and of 200\,K for the emission, 
in agreement with the estimates obtained from CO (Sect.\,\ref{rotco}).
 The central velocity and the line width of each component are free parameters of the fit, and the results are given in Table\,\ref{oh_para}. Although only part of the $^2\Pi_{3/2}, J=5/2 \rightarrow 3/2$ triplet at 2509.9\,GHz falls in the observed spectrum, the blue-shifted wing of this feature was used to constrain the fit. Similarly, the red-shifted emission was also included in the analysis; however, the velocity range ([-50,-2]\,km\,s$^{-1}$) used to derive the OH column density in the blue-shifted lobe of the outflow is not influenced by the results of the red-shifted wing.  
 The best fit of the OH ground-state triplet is shown in Fig.\,\ref{oh_fit}. 

\begin{table}
\caption[]{Fit results for the OH $^2\Pi_{3/2}, J=5/2 \rightarrow 3/2$ triplet at 2514\,GHz.}\label{oh_para}
\begin{center}
\begin{tabular}{cccccc}
 \hline  \hline
&$\varv$&$\Delta \varv$&$N_{\rm{OH}}$&$T_{\rm{ex}}$&Em./Abs.\tablefootmark{a}\\
&(km\,s$^{-1}$)&(km\,s$^{-1}$)&($10^{14}$cm$^{-2}$)&(K)\\
\hline

1 & -13.2& 35.2 &  4.1 &     10 &    A\\
2 & +7.3 & 13.5 &  5.6 &     10 &    A\\
3 & +20.2& 3 4.9 & 5.4  &    10 &    A\\
4 & +34.3&  15.3 & 0.2 &     200 &    E\\
5 & +55.1&  30.6 & 0.2 &     200 &    E\\

\hline
\end{tabular}
\end{center}
\tablefoot{\tablefoottext{a}{E is for the component in emission, A for those in absorption.}}
\end{table}

\subsection{Abundances}

In Table\,\ref{abu} we report abundances of OH, H$_2$O, and HF relative
to CO for the blue- ($[-50,-2]$\,km\,s$^{-1}$) and the red-shifted
($[+32,+70]$\,km\,s$^{-1}$) velocity ranges.  Given the similarities
of the line profiles, we assumed that all molecules trace the same
gas. 
 In the blue-shifted wing the water abundance relative to
 CO ranges between $2\times10^{-1}$ and $4\times10^{-2}$.
Since  \citet{gusdorf15} demonstrated that [CII] is the main form of carbon in G5.89--0.39 in the warm component,
while CO and [CI] have similar column densities, we 
assumed that the  total (CO+[CI]+C[II]) abundance of carbon relative to H$_2$
is $1.4\times 10^{-4}$ and that the column density of [CI] is equal to that of the warm CO \citep{gusdorf15}. This results
in H$_2$O/H$_2$ abundances of
$10^{-7}-10^{-6}$ in agreement with determinations in outflows
from low-mass YSOs 
\citep[e.g., ][]{2010A&A...518L.113L,2012A&A...542A...8K} but
lower than in molecular outflows from massive YSOs \citep{2010A&A...521L..28E,2014A&A...564L..11L}. The water abundance in the red-shifted lobe is one order of magnitude smaller than at blue-shifted HV (H$_2$O/CO$\sim5\times 10^{-3}$, H$_2$O/H$_2$ $\sim 3\times 10^{-8}$). 
For OH, we estimate
an abundance relative to CO of $(3-6)\times 10^{-3}$, and of $(1-3)\times10^{-8}$ relative to H$_2$.
The resulting OH to H$_2$O abundance is 0.01 for the blue-shifted lobe, and
0.6 in the red-shifted velocity range. Given the large uncertainties in the estimates of the OH and H$_2$O column densities, these abundances are
consistent with models of fast dissociative J-type shocks \citep{1989ApJ...344..251N} and of slower UV-irradiated C-type shocks \citep{1999ApJ...525L.101W}.
For HF, we derive an abundance relative to CO of $2\times10^{-3}$
and of $10^{-8}$ relative to H$_2$.
While 
in the diffuse medium HF is an excellent tracer of molecular hydrogen
\citep[e.g.][]{2010A&A...518L.108N,2011ApJ...742L..21M} with a
relative abundance to H$_2$ of $\sim10^{-8}$
\citep{2013ApJ...764..188I}, the abundance of HF
relative to H$_2$ is not well constrained in dense environments, where
depletion of F on
grain surfaces is believed to play an important role
\citep{2010A&A...518L.109P}.  Its abundance in molecular outflows from massive YSOs is even more uncertain:
while
\citet{2010A&A...518L.109P} derived a lower limit to HF/H$_2$ of
$1.6\,10^{-10}$ in Orion-KL, \citet{2012ApJ...756..136E} found $3.6\,10^{-8}$ in AFGL\,2591.
Our measurement of the HF abundance in a molecular outflow from a massive YSO 
falls in between the two previous estimates.

\section{Mass-loss rate and energetics of the outflow system}\label{sec_massloss}

The ${\rm[OI]_{63\mu m}}$ line is expected to be the
brightest atomic line in jets from YSOs because of the high abundance of atomic oxygen and favourable excitation conditions \citep[e.g., ][]{1989ApJ...342..306H}.
For temperatures below 5000\,K, \citet{1985Icar...61...36H} suggested that 
the luminosity of the 
 ${\rm[OI]_{63\mu m}}$  line is directly proportional to the mass-loss rate from a protostar:

\begin{equation}\label{hollenbach}
\frac{\dot{M}}{M_\odot \mathrm{yr}^{-1}}= 10^{-4}\frac{L_{\rm[OI]\,63\mu m}}{L_\odot}
.\end{equation}

To allow the comparison with the estimates of the outflow energetics from CO, [CI]
and [CII] from \citet{gusdorf15}, we smoothed the [OI]$_{\rm{63\mu m}}$ data to a
resolution of 12\farcs5. The corresponding
[OI]$_{\rm{63\mu m}}$ luminosities for the blue-  and red-shifted  wings are 4.8\,$L_\odot$ and 1.9\,$L_\odot$
(note that in Table\,\ref{lfir} $L_{\rm{OI}}$ are reported on  a 14\farcs5 scale).
We assumed that contamination of PDR emission at these velocities is negligible since the
velocity range used for the integration starts at $|v-v_{\rm {LSR}}|> 12$\,km\,s$^{-1}$.
The mass-loss rate estimates derived from Eq.\,\ref{hollenbach} for the blue- and
the red-shifted wings of [OI]$_{\rm{63\mu m}}$ are reported in Table\,\ref{mdot}
together with the values from \citet{gusdorf15} based on CO, [CI] and
[CII].

Equation\,\ref{hollenbach} relies only on the assumption that [OI]$_{\rm{63\mu m}}$ is the dominant coolant of the gas in
shocks with $T<5000$\,\,K. This assumption was verified to be valid for pre-shock densities $n<10^6$\,cm$^{-3}$ by
\citet{1989ApJ...342..306H}. However,  the shock models of \citet{1989ApJ...342..306H} most likely do not apply to massive star-forming regions
since the UV radiation field from the protostar is completely neglected and preshock densities may be higher.
Indeed, \citet{gusdorf15} verified that
these models fail to reproduce the [CII] emission in G5.89--0.39 very likely because the only possible cause
of dissociation is the UV field generated by the shock propagation
itself. Moreover, the models of \citet{1989ApJ...342..306H} assumed a weak magnetic field, which is in contradiction with
observations \citep{2009ApJ...695.1399T}.
Therefore, we also computed the mass-loss rate of the jet system in G5.89--0.39 directly as

\begin{equation}\label{mdot_td}
\dot{M} = \frac{M}{t_d} 
,\end{equation}

where $t_d$ is the dynamical time of the outflow, and $M$ the total mass in the outflow

\begin{equation}\label{td}
t_d = \frac{R}{\Delta\varv_{\rm{max}}} 
\end{equation}

\begin{equation}\label{m_oi}
M = N\times\pi R^2 
.\end{equation}

In Eqs.\,\ref{td} and \,\ref{m_oi}, $R$ is radius of each lobe of the
outflow assuming a circular size with a diameter of 6\farcs6 because the [OI]$_{\rm{63\mu m}}$ emission is
mostly compact,
$\Delta\varv_{\rm{max}}$ is the zero-intensity line width of the [OI]
63\,$\mu$m line ($\Delta\varv_{\rm{max_{blue}}}$=50\,km\,s$^{-1}$ and
$\Delta\varv_{\rm{max_{red}}}$=58\,km\,s$^{-1}$ assuming a rest
velocity of 10\,km\,s$^{-1}$). The dynamical timescale of the [OI]
outflow system is 350\,yr for the red lobe and 400\,yr for the
blue lobe, in agreement with the estimate of \citet{gusdorf15} of
380\,yr for both lobes based on CO.  In Eq.\,\ref{m_oi}, $M$ is the
total mass of the outflow and is derived over a 12\farcs5 beam;
$N$ is the total column density of
H$_2$ at high velocities and is derived from the column density of
[OI] assuming a relative abundance to H$_2$.  Our estimate of the mass of the red-shifted wing
(and of the other parameters related to the energetics) is probably a lower limit since the [OI] column density
at these velocities is determined under the assumption that the emission is optically thin, while the line profile
suggests high opacities.

Gas-phase oxygen abundances between $3.07\times 10^{-4}$ and $4.85\times 10^{-4}$ are reported in
the literature \citep[e.g., ][]{2015ApJ...801..110P,2006ApJ...639..929B}. Therefore, we adopted
an oxygen abundance of $4\times 10^{-4}$ and assumed 
that all oxygen is in atomic form and in CO (whose column density is 1--4\,\% of that of [OI], while those of H$_2$O and OH are below 0.1\,\%). The corresponding  total column densities are
$8.3\times 10^{21}$\,cm$^{-2}$ and  $7.5\times 10^{21}$\,cm$^{-2}$  for the blue- and red-shifted lobes, respectively. The
masses are 0.1\,$M_\odot$ for both lobes, and 
 mass-loss rates  $3.1\times 10^{-4} \rm{M_\odot\,yr^{-1}}$ and $3.2\times 10^{-4} \rm{M_\odot\,yr^{-1}}$ for the blue- and red-shifted lobes,
in good agreement with the estimates based on Eq.\,\ref{hollenbach} and on those based 
on CO, [CI], and [CII] from \citet{gusdorf15}. This suggests that the assumption behind Eq.\,\ref{hollenbach} (that the [OI]$_{\rm{63\mu m}}$ is the main coolant at high velocities) also holds
in massive jet systems. In Sect.\,\ref{sec_fir} we investigate the contribution of several species to the
far-IR line luminosity in different velocity ranges to confirm that [OI]$_{\rm{63\mu m}}$ is the main coolant at high velocities.

\begin{table*}
\caption[]{[OI]$_{\rm{63\mu m}}$ mass-loss rates and comparison with estimates from other tracers.}\label{mdot}
\begin{center}
\begin{tabular}{lccc}
 \hline  \hline
 &$\dot{M}\tablefootmark{a}_{{\rm{shock [OI]}}}$&$\dot{M}\tablefootmark{b}_{\rm{[OI]}}$&$\dot{M}\tablefootmark{c}$\\
 &  $(10^{-4} \rm{M_\odot\,yr^{-1}})$&$(10^{-4} \rm{M_\odot\,yr^{-1}})$&$(10^{-4} \rm{M_\odot\,yr^{-1}})$\\
\hline
Blue ($[-50,-2]$\,km\,s$^{-1}$)&4.8&>3.1&$>(2.1-2.2)$\\
Red  ($[+32,+65]$\,km\,s$^{-1}$)&1.9&>3.2&$>(0.9-1.1)$\\

\hline
\end{tabular}
\end{center}
\tablefoot{\tablefoottext{a}{$\dot{M}_{{\rm{shock}[OI]}}$ is the mass-loss rate based on the formula from
  \citet{1985Icar...61...36H};}
 \tablefoottext{b}{$\dot{M}_{\rm{[OI]}}$ is the mass-loss rate based on Eq.\,\ref{mdot_td};}
  \tablefoottext{c}{Mass-loss rate estimate from \citet{gusdorf15} for the warm component based on CO, [CI] and [CII]} in the velocity ranges  $[-75,+4]$\,km\,s$^{-1}$ for the blue-shifted lobe and $[+16.5,-75]$\,km\,s$^{-1}$ for the red-shifted emission.}
\end{table*}

The energetic of the jet system can be further characterised by estimating the
momentum $P$, the mechanical force $F_m$,
the kinetic energy $E_k,$ and the mechanical luminosity $L_{m}$ through

\begin{equation}\label{p}
P = M\times{\Delta\varv_{\rm{max}}} 
,\end{equation}

\begin{equation}\label{fm}
F_m = M\times{\Delta\varv_{\rm{max}}}/t_d
,\end{equation}

\begin{equation}\label{ek}
E_k = M\times{\Delta\varv_{\rm{max}}}^2/2
, \text{and}\end{equation}

\begin{equation}\label{lm}
L_{m} = \frac{E_k}{t_d}
.\end{equation}

Results are summarised in Table\,\ref{para}.  
The parameters related to the energetics of the outflow
agree well with  those estimated by
\citet{gusdorf15}. However, our estimates are
lower limits: no corrections for beam dilution or for the
opacity of the [OI]$_{\rm{63\mu m}}$ line were implemented in the calculation of the [OI]
column density at blue-shifted velocities, and we have no estimates for the $N_{\rm{[OI]}}$ column density
in the red-shifted wing. Given these limits and the similar mass-loss rates found for the atomic jet and the molecular outflow, our results suggest that the molecular
outflow system in the region is driven by the  jet traced by [OI]$_{\rm{63\mu m}}$.

\begin{table*}
\caption[]{Jet parameters based on the [OI]$_{\rm{63\mu m}}$ emission and comparison with outflow parameters from CO, [CI], and [CII] from \citet{gusdorf15} for the warm component. Results are obtained assuming an oxygen abundance  relative to H of $4\times10^{-4}$ and a carbon abundance of $1.4\times10^{-4}$.}\label{para}
\begin{center}
\begin{tabular}{lcccc}
  \hline  \hline
  &\multicolumn{2}{c}{[OI]$_{\rm{63\mu m}}$}&\multicolumn{2}{c}{CO+[CI]+[CII]}\\
  & Blue&Red& Blue&Red\\
  \hline
 $N\,(10^{21}$\,cm$^{-2}$)&8.3&>7.5&4.2--4.6&>1.9--2.0\\
 $M\, (M_\odot)$& 0.1&>0.1&0.2&>0.1\\
 $\Delta\varv_{\rm{max}}$\,(km\,s$^{-1})$&50&58&--\tablefootmark{a}&--\tablefootmark{a}\\
 $t_d$\,(yr)&400&350&--\tablefootmark{a}&--\tablefootmark{a}\\
 $P\,(M_\odot$\,km\,s$^{-1})$&6.1&>6.5&8--9&>3.5--4.0\\
 $F_m\,(10^{-2}\,M_\odot$\,km\,s$^{-1}$\,yr$^{-1}$)&2&>2&1.1&>0.5xs\\
 $E_k\,(10^{45}$\,erg)&3.1&>3.7&4.0--4.3&>1.8--2.0\\
 $L_{m}\,(L_\odot)$&63&>89&42--45&>19--21\\
\hline
\end{tabular}
\end{center}
\tablefoot{\tablefoottext{a}{See \citet{gusdorf15} for values for single species.}}
\end{table*}

\section{Far-infrared line luminosity}\label{sec_fir}
Far-infrared atomic and molecular features are commonly used to derive
the gas cooling-budget in star-forming regions
\citep[e.g.,][]{2002ApJ...574..246N,2013A&A...552A.141K,2014A&A...562A..45K}. Following
\citet{2013A&A...552A.141K,2014A&A...562A..45K}, we define the total far-infrared line cooling
($L_{\rm{FIRL}}$) as the sum of the line emission luminosity from the
atomic gas (from [OI] and [CII] in our case, $L_{\rm{[OI]}}$ and
$L_{\rm{[CII]}}$ , respectively) and from the molecular gas (H$_2$O, CO
and OH, $L_{\rm{[H_2O]}}$, $L_{\rm{[CO]}}$, and $L_{\rm{[OH]}}$). 
In our analysis, we considered only spectral features with frequencies higher than 1000\,GHz.
When
possible, we computed the luminosity of each species on datasets
deconvolved to 14\farcs5, the spatial resolution of the CO(16--15)
data. OH and H$_2$O observations are single pointings, but with  resolutions  similar to that of CO(16--15) (15\farcs3 and
11\farcs6 for OH at 1834--1837\,GHz and 2514\,GHz; 12\farcs7 and
19\farcs3 for H$_2$O at 1661-1669\,GHz and at 1113\,GHz). The far-infrared line luminosity
calculated in this way is a lower limit to its real value since it is
based only on the spectral features presented in this paper. Table\,\ref{lfir} summarises our results
together with estimates from \citet{2014A&A...562A..45K}, who analysed G5.89--0.39 in their sample of
massive star-forming regions.
 Based on PACS observations of CO, H$_2$O, OH, [OI], they derived
a far-infrared line luminosity of 8.8\,$L_\odot$ for G5.89--0.39.
The CO(16--15) transition is the
strongest CO line in the PACS range and contributes to $\sim13$\% to
the total CO luminosity. We find a CO(16--15) line luminosity integrated over the whole line profile of 0.65\,$L_\odot$
, which agrees well with the results of \citet{2014A&A...562A..45K} ($L_{\rm{CO(16-15)}}\sim0.5\,L_\odot$) if we consider that our beam is larger than
the spaxel size (9\farcs4) and that the CO emission is extended \citep{gusdorf15}.
For the atomic gas,
$L_{\rm{[CII]}}$ was not analysed by \citet{2014A&A...562A..45K} because of
contamination from the off position. The  $L_{\rm{[OI]}}$ from \citet{2014A&A...562A..45K} reported in Table\,\ref{lfir}
is
only based on the 63\,$\mu$m flux as in our study. We note that the 145\,$\mu$m [OI] line does not contribute significantly to the total [OI] luminosity  \citep{2014A&A...562A..45K}. Our higher
$L_{\rm{[OI_{63{\mu}m}]}}$ (5.7\,$L_\odot$) compared to \citet{2014A&A...562A..45K} (3.7\,$L_\odot$)
is due to the larger beam used in our analysis. Indeed, on a 6\farcs6
resolution, $L_{\rm{[OI_{63{\mu}m}]}}$ is 3.4\,$L_\odot$ , which agrees well  with
previous estimates.
For the other species, a direct comparison between our results and those presented by \citet{2014A&A...562A..45K}
is not straightforward.
Values for single water lines
are not reported in \citet{2014A&A...562A..45K}, and the OH luminosity ($L_{\rm{OH}}=0.5\,L_\odot$) is calculated from the
triplets at 71\,$\mu$m and 163\,$\mu$m, while our estimate (0.44\,$L_\odot$) is based on
the two triplets at 163\,$\mu$m and on the ground-state lines at
119\,$\mu$m at red-shifted high velocities. However, our findings generally agree well with  \citet{2014A&A...562A..45K}. On the spatial scale considered in our analysis ($\sim20\,000$\,AU),
[OI] and CO (considering the whole far-infrared ladder from \citet{2014A&A...562A..45K}) are the main
contributors to the cooling.  Water is 1\% of $L_{\rm{FIRL}}$ if we consider only
the lines presented in this paper, and it accounts for up to 10\% of the
total cooling if we include the results of
\citet{2014A&A...562A..45K}. Therefore, water is a minor contributor to the total far-infrared line luminosity in this source.
Finally, [CII] amounts to $\sim 4\%$ of
the total line luminosity reported by \citet{2014A&A...562A..45K}.

Given the high spectral resolution of all our datasets, we can
determine the contribution of each species to the total cooling budget 
in different velocity regimes (see Table\,\ref{lfir}).
While the total [OI]$_{\rm{63\mu m}}$ luminosity is dominated by emission at
blue-shifted velocities, the CO(16--15) luminosity arises almost
equally from the blue- and red-shifted wings and from ambient velocities.  If
we assume that each of the three velocity components identified in CO
(blue- and red-shifted emission, ambient emission) contributes to a
third of the CO luminosity reported by \citet{2014A&A...562A..45K}
over the whole ladder \citep[as suggested
by the CO(6--5), (7--6) and (16--15) lines, see also][]{gusdorf15}, the
far-infrared line cooling budget of the outflowing gas is dominated by
atomic oxygen at blue-shifted velocities. In the red-shifted wing,
$L_{\rm{[OI]}}$ and $L_{\rm{[CO]}}$ are comparable. At ambient
velocities, the far-infrared line cooling budget is dominated by CO as all
other species are seen in absorption in this velocity range. The three
water lines analysed in this study have a line luminosity of
0.1\,$L_\odot$ in the red-shifted wing, which makes water also a minor contributor to the total cooling
in the outflowing gas.

\begin{table*}
\caption{Far-infrared line luminosities.}\label{lfir}
\begin{center}
\begin{tabular}{lcccccc}
  \hline  \hline
  \multicolumn{7}{c}{This work (14\farcs5 beam)}\\
  \hline
Velocity range &$L_{\rm{CO(16-15)}}$
               &$L_{\rm{OH}}\tablefootmark{a}$
               &$L_{\rm{H_2O}}\tablefootmark{b}$
               &$L_{\rm{OI\,63\mu m}}$
               &$L_{\rm{CII}}$
               &$L_{\rm{FIRL}}$\\
&$(L_\odot)$&$(L_\odot)$&$(L_\odot)$&$(L_\odot)$&$(L_\odot)$&$(L_\odot)$\\
\hline 
total profile ($[-50,+65]$\,km\,s$^{-1}$)&$0.65$&$0.44$ &--                  & $5.7$ & $0.42$&7.21\\
HV-red  ($[+47,+65]$\,km\,s$^{-1}$)      &--    &$0.08$ &$0.03$                 & $0.9$& $0.02$&1.03\\ 
LV-red  ($[+32,+47]$\,km\,s$^{-1}$)      &$0.06$&$0.13$ &$0.09$                 & $1.2$& $0.06$&1.48\\
ambient\tablefootmark{c} ($[-2,+26]$\,km\,s$^{-1}$)      &$0.42$&$0.12$ &$0.08\tablefootmark{d}$&  --   & $0.1$&0.72\\
HV-blue ($[-35,-50]$\,km\,s$^{-1}$)      &--           &-- &--                  &$0.02$ & --           &0.02\\
LV-blue ($[-35,-2]$\,km\,s$^{-1}$)      &$0.17$&-- &--                         &$5.3$ &$0.2$ &5.67\\
 \hline
 \multicolumn{7}{c}{Values from \citet{2014A&A...562A..45K} (9\farcs4 beam)}\\
 \hline
Velocity range &$L_{\rm{CO}}\tablefootmark{e}$
               &$L_{\rm{OH}}\tablefootmark{f}$
               &$L_{\rm{H_2O}}\tablefootmark{f}$
               &$L_{\rm{OI\,63\mu m}}\tablefootmark{g}$\\

 total profile &$3.9$&$0.5$ &$0.8$& $3.7$ & --&8.8\\ 
  \hline

\end{tabular}
\end{center}
\tablefoot{For all measurements, error bars are dominated by calibration uncertainties (20\,\%). 
  \tablefoottext{a}{$L_{\rm{OH}}=L_{\rm{1835GHz}}+L_{\rm{1838GHz}}+L_{\rm{2514GHz}}$. Data have a spatial resolution of 15\farcs3 (1835\,GHz and 1838\,GHz) and 11\farcs3 (2514\,GHz)}
\tablefoottext{b}{$L_{\rm{H_2O}}=L_{\rm{1113GHz}}+L_{\rm{1661GHz}}+L_{\rm{1669GHz}}$. Data have a spatial resolution of 19\farcs0 (1113\,GHz) and 12\farcs7 (1661 and 1669\,GHz).}
\tablefoottext{c}{The velocity range $[+28,+32]$\,km\,s$^{-1}$ is not taken into account because of contamination from a telluric line in the CO(16--15) spectrum (Sect.\,\ref{sec_cal}).}
\tablefoottext{d}{Computed only on the 1661\,GHz line.}
\tablefoottext{e}{Computed over the whole CO from $J=14-13$ to $J=46-45$.}
\tablefoottext{f}{Computed over all OH and H$_2$O lines detected in emission in the PACS spectral range.}
\tablefoottext{g}{Computed on the ${\rm[OI]_{63\mu m}}$ line.}
}

\end{table*}

\section{Discussion}

\subsection{Atomic and molecular gas in jet/outflow systems}\label{dis_oi}
In Sect.\,\ref{sec_massloss} we found that the mass-loss rates
estimated in G5.89--0.39 from [OI]$_{\rm{63\mu m}}$ and from molecular gas are similar.
The [OI]$_{\rm{63\mu m}}$ mass-loss rate was estimated at high velocities ($|v-v_{\rm {LSR}}|> 12$\,km\,s$^{-1}$) assuming that PDR contamination is negligible \citep[see][for typical line-widths in PDRs for different tracers]{2009A&A...498..161V,2012A&A...544A.110P}. 
This suggests that the atomic jet in G5.89--0.39 is powerful enough to
drive the larger scale molecular outflow(s).  Similar comparisons
between mass-loss rates from jets and molecular outflows were
attempted in a few other massive YSOs with different jet tracers (SiO,
H$_2$, ionised gas). In IRAS\,20126+4104 and IRAS\,18151--1208, the
jet seems to be driving the corresponding outflow
(\citealt{1999A&A...345..949C,2000ApJ...535..833S,2008A&A...485..137C}
for IRAS\,20126+4104;
\citealt{2002A&A...383..892B,2004A&A...425..981D} for
IRAS\,18151–-1208), while in Ceph\,A and G192.16-3.82 the observed
jets do not have enough momentum to support the outflows
\citep{1999ApJ...523..690S,1994ApJ...430L..65R,1999ApJ...514..287G}.

The comparison of mass-loss rates derived from the atomic gas with
molecular estimates also gives important constraints on the role of
the atomic and  molecular components in outflow/jet systems.
Pioneering studies with ISO of outflows from low-mass YSOs
compared mass-loss rates from CO with estimates from the
${\rm[OI]_{63\mu m}}$ line
\citep{1997ApJ...476..771C,2001ApJ...555...40G} and found that the two
values agree fairly well despite the crude assumptions in the analysis
of the [OI]$_{\rm{63\mu m}}$ data.  Recently, {\it Herschel} improved the situation by
increasing the spatial resolution of ${\rm[OI]_{63\mu m}}$
observations by a factor of ten. \citet{nisini2015} analysed
five jets from Class 0 low-mass YSOs and found that only in two sources
the mass-loss rate from [OI]$_{\rm{63\mu m}}$ is comparable with estimates based on CO,
while in the other three cases $\dot{M}_{{\rm[OI]}}$ is more than an
order of magnitude smaller than the corresponding
$\dot{M}_{{\rm[CO]}}$. The authors suggested that their results fit in
an emerging scenario where jets from low-mass YSOs undergo an
evolution in their composition with time:  jets in the youngest sources
would be mainly molecular, while the atomic component would become
progressively more dominant as the jet gas excitation and ionisation
increases \citep[see also][]{2012A&A...545A..44P} and would be dominant only in an intermediate phase of jet evolution. Our finding that
the mass-loss rate from [OI]$_{\rm{63\mu m}}$ is comparable with that from the
molecular gas, and therefore that the atomic jet has an important role
in driving the corresponding molecular outflow, could fit in this
scenario since G5.89--0.39 already hosts an
ultra-compact H{\sc II} region. However, the statistics on [OI]$_{\rm{63\mu m}}$ in
jets is too small to reach any solid conclusion: only a tenth of jets
from low-mass protostars were mapped in [OI]$_{\rm{63\mu m}}$
\citep{2012A&A...545A..44P,2012A&A...548A..77G,nisini2015}, and
G5.89--0.39 is the first massive YSO observed in [OI]$_{\rm{63\mu m}}$
with proper spectroscopic resolution to distinguish the contribution
of high-velocity emission from ambient gas.

\subsection{Quest for high-spectral resolution}

Figure\,\ref{spectra} shows that  the [OI]$_{\rm{63\mu m}}$ spectra in G5.89--0.39 are
dominated by emission from the jet system, while the low-velocity
range, most likely associated with PDR emission from the gas surrounding
the ultra-compact H{\sc ii} region, is severely contaminated by
absorption features. Thus in G5.89--0.39, the PDR contribution to the [OI]$_{\rm{63\mu m}}$ line is completely
missing in the data.

Previous observations at lower spectral resolution \citep[see for
  example][]{1996ApJ...462L..43P,2006A&A...446..561L,2015ApJ...801...72R}
with KAO, ISO and {\it Herschel} suggested that the
[OI]$_{63\mu{\rm{m}}}$ transition may be considerably affected by
absorption from foreground clouds and from the source envelope in
different environments. These findings question the choice of
[OI]$_{63\mu{\rm{m}}}$ as a tracer of PDRs and of star-formation rates in external galaxies in
case of spectrally unresolved lines.  The case of G5.89--0.39 is
exceptionally striking since {\it Herschel} PACS observations of the
same region show a perfect Gaussian profile for the
[OI]$_{63\mu{\rm{m}}}$ with no hint of absorption
\citep[][ their Fig.\,2]{2014A&A...562A..45K} with a spectral resolution of
90\,km\,s$^{-1}$.  The importance of absorptions in the line depends
on the total column density of the source itself and on the position
of the source in the Galaxy, which determines the number of
intervening interstellar clouds along the line of sight to the source
itself. Therefore,  high-spectral
resolution is in general a strong requirement for this kind of
studies.

\section{Conclusion}

We reported the first high spectral resolution observations of the
[OI] line at 63\,$\mu$m.
The target of our study was the inner part of the massive star-forming region G5.89--0.39,
an ideal source to investigate the contribution of PDR emission, absorption features, and
jet emission at high velocities since it hosts an ultra-compact H{\sc ii} region and
at least three outflows. We complemented the SOFIA [OI]$_{\rm{63\mu m}}$ observations with spectroscopically resolved
data from {\it Herschel} and APEX of the other major coolants of the gas (CO, OH, H$_2$O, [CII]) to study their contributions 
to the total far-IR line luminosity in different velocity ranges. Our main results  can be summarised as follows:

\begin{itemize}
\item The [OI]$_{\rm{63\mu m}}$ spectra are severely affected by
  absorption from the source envelope and from intervening interstellar
  clouds along the line of sight. Emission is detected at
  high velocities.  It is compact and associated with the molecular
  outflows detected along the north-south direction in previous
  observations of CO.
\item The parameters of the jet system derived from the
  [OI]$_{\rm{63\mu m}}$ agree well with those estimated for
  the warm molecular outflow system associated with the [OI] emission. This
  suggests that at least in this source the molecular outflow is driven
  by the atomic jet seen in [OI].
\item CO and [OI] contribute in a similar way to the total far-IR line luminosity
  of G5.89--0.39 when the full velocity range is considered. However,
  the [OI] emission is dominated by the high-velocity range, and indeed [OI] is the main
  contributor to the cooling budget of the gas at high velocities.
\item The line luminosity of the [OI] line at high velocities can be used as tracer of the mass-loss rate
  of the jet since [OI] is the main coolant of the gas in this velocity regime.
\end{itemize}

Our [OI]$_{\rm{63\mu m}}$ observations clearly demonstrate that the 63\,$\mu$m line can
be heavily contaminated by absorption features from different clouds
along the line of sight and that its line luminosity can be
completely associated with jets in extreme sources such as G5.89--0.39. Therefore,
its use as PDR and star-formation rate tracer might be limited if the
profile of the line is not resolved. Velocity-resolved observations are now made be possible by the
SOFIA telescope
with the GREAT spectrometer and its upcoming high-frequency array
upGREAT.

\begin{acknowledgements}
  We like to thank an anonymous referee and Malcolm Walmsley for comments and suggestions that improved the clarity of the paper. 
  We are grateful to Brunella Nisini and Linda Podio for
  helpful discussions 
  and for allowing us access to their results before publication.
This work is based in part on observations made with the NASA/DLR Stratospheric Observatory for Infrared Astronomy (SOFIA).
SOFIA is jointly operated by the Universities Space Research Association, Inc. (USRA), under NASA contract NAS2-97001, and the Deutsches SOFIA Institut (DSI) under DLR contract 50 OK 0901 to the University of Stuttgart.
\end{acknowledgements}

\Online

\begin{appendix}
\section{Correction for atmospheric absorption by mesospheric oxygen}\label{cal}
\begin{figure}
\includegraphics[width=\columnwidth]{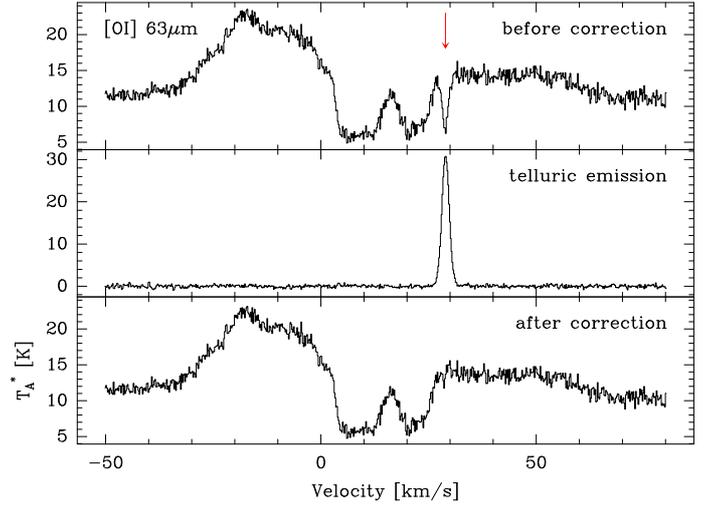}
\caption{Top: $63.2\mu$m [OI] line of G5.89--0.39 after correction for 
the broad atmospheric absorption, but with the telluric [OI] line still
uncorrected. Centre: Telluric [OI] emission line. Bottom: [OI] spectrum
after the correction derived from a Gaussian fit to the telluric line.
The continuum level is not corrected for the fact that GREAT operates in double-sideband.}
\label{app:1}
\end{figure}

The correction for atmospheric absorption of the spectra obtained with GREAT 
follows the usual procedure described in \citet{2012A&A...542L...4G}. This
concerns absorption due to water vapour and the quasi-continuous,
collision-induced absorption by O$_2$ and N$_2$. For the telluric [OI] line
at $63.2~\mu$m only few spectra with high resolution exist.
Since the width of the telluric [OI] line is $\sim 1$\,kms$^{-1}$,
the corresponding spectral channels could
simply be masked \citep[e.g.,][]{1996ApJ...464L..83B}. Here we followed a different approach.
 Since the telluric [OI] line originates
in the mesosphere, we can approximate its radiative transfer by a single
layer with a constant source function and absorption coefficient, that is,
\begin{equation}
T_{\rm L, \nu} = T_{\rm atm, \nu} \eta_{\rm f} \left(1-exp{(-\tau_\nu)}\right)
,\end{equation}
where $T_{\rm L, \nu}$ and $T_{\rm atm, \nu}$ are the Rayleigh-Jeans
equivalent values for the line temperature and mesopheric temperature,
respectively. $\eta_{\rm f} = 0.97$ is the forward efficiency of GREAT. This
expression holds only for the [OI] line, contributions from the other
atmospheric species are omitted because the standard calibration procedure
already accounts for them. The transmission in the [OI] line, along the
sightline through the stratosphere, thus becomes
\begin{equation}
exp{(-\tau_\nu)} = 1-\frac{T_{\rm L, \nu}}{T_{\rm atm, \nu}\eta_{\rm f}}\,.
\end{equation}
Because of the mesospheric origin, the telluric [OI] line has a Gaussian
profile, 
\begin{equation}
  T_{\rm L, \nu} = T_{0, \nu} \cdot exp{\left(-4\ln{2}\frac{(\nu-\nu_0)^2}{\Delta\nu^2}\right)}
,\end{equation}

where $\nu_0 = 4744.77749$~GHz is the line centre frequency, and $\Delta\nu$ its
full width at half-maximum. We obtain $T_{0, \nu}$ and $\Delta\nu$ from a
least-squares fit to the line profile $T_{\rm L, \nu}$. The Rayleigh-Jeans
equivalent atmospheric temperature at $\nu_0$ is kept as a free parameter and
adjusted in such a way that the extinction correction $\exp{(\tau_\nu)}$
removes the telluric absorption in front of the [OI] line from G$5.89-0.39$.
The only prerequesite needed here is that the spectrum be free from
narrow components within $\sim 2\Delta\nu$ from $\nu_0$, which is the case
here. These calibration steps are shown in Fig.~\ref{app:1}.

\end{appendix}

\end{document}